\documentclass[12pt,preprint,letterpaper]{aastex}
\usepackage{graphicx}

\begin{document}

\title{New Approaches To Photometric Redshift Prediction Via Gaussian Process
Regression In The Sloan Digital Sky Survey} 

\author{M. J. Way\altaffilmark{1,2}}
\affil{NASA Goddard Institute for Space Studies,
2880 Broadway, New York, NY 10029, USA}
\author{L. V. Foster}
\affil{Department of Mathematics, San Jose State University,
One Washington Square, San Jose, CA 95192, USA}
\author{P. R. Gazis}
\affil{NASA Ames Research Center, Space Sciences Division,
MS 245-6, Moffett Field, CA 94035, USA}
\author{A. N. Srivastava}
\affil{NASA Ames Research Center, Intelligent Systems Division,
MS 269-4, Moffett Field, CA 94035, USA}

\altaffiltext{1}{NASA Ames Research Center, Space Sciences Division,
MS 245-6, Moffett Field, CA 94035, USA}
\altaffiltext{2}{Department of Physics and Astronomy,
Uppsala, Sweden}

\begin{abstract}
Expanding upon the work of \cite{Way06} we demonstrate how the use of training
sets of comparable size continue to make Gaussian process regression (GPR) a
competitive approach to that of neural networks and other least-squares
fitting methods.  This is possible via new large size matrix inversion
techniques developed for Gaussian processes (GPs) that do not require that the
kernel matrix be sparse. This development, combined with a neural-network
kernel function appears to give superior results for this problem. Our best
fit results for the Sloan Digital Sky Survey (SDSS) Main Galaxy Sample using
$u$,$g$,$r$,$i$,$z$ filters gives an rms error of 0.0201 while our results
for the same filters in the luminous red galaxy sample yield 0.0220.
We also demonstrate that there appears to be a minimum number of training-set
galaxies needed to obtain the optimal fit when using our GPR
rank-reduction methods. We find that morphological
information included with many photometric surveys appears, for the most
part, to make the photometric redshift evaluation slightly worse
rather than better.  This would indicate that most morphological information
simply adds noise from the GP point of view in the data used herein.
In addition, we show that cross-match catalog results involving combinations of
the Two Micron All Sky Survey, SDSS, and Galaxy Evolution
Explorer have to be evaluated in the context of the resulting cross-match
magnitude and redshift distribution. Otherwise one may be misled into overly
optimistic conclusions.
 
\end{abstract}


%%%%%%%%%%%%%%%%%%%%%%%%%%%%%%%%%%%%%%%%%%%%%%%%%%%%%%%%%%%%%%%%%%%%%%%

\section{Introduction}\label{sec:intro}

General approaches to calculating photometric redshifts from broad band
photometric data have been discussed elsewhere recently
\citep[][hereafter Paper I]{Way06}.  These involve template based approaches
and what are referred to as training-set approaches. In this paper we expand
upon the training-set approaches outlined in Paper I using Gaussian processes
(GPs).  Previously we were limited to training set sizes of order 1000 because
a matrix inversion of order 1000$\times$1000 was required for calculating
the GPs. Part of the limitation was due to the amount of single thread
accessible RAM on our circa 2005 32bit computers, meaning that one could
not invert a matrix larger than about $O$(1000$\times$1000) in size at one time
within Matlab\footnote{http://www.mathworks.com}, our choice for implementing
GPs. Today one can now use commodity based 64bit workstations and invert
matrices of $O$(20000) within Matlab. However, even this is a small fraction
of the total potential size of today's photometric redshift training sets.
For this reason we have developed new non-sparse rank-reduction matrix inversion
techniques that allow one to use over 100,000 training samples.  From this
work we demonstrate that the new rank-reduction methods only require
approximately 30-40,000 samples to get the optimal possible fit from GPs on
Sloan Digital Sky Survey \citep[][SDSS]{York2000} data.

Since Paper I several new approaches to Galaxy photometric redshifts
from broad band photometry have come about along with expansion and refinement
of previously published methods. Below is a summary of some of these
approaches.

\cite{Kurtz07} have used
the Tolman surface brightness test ($\mu$-- Photo$Z$) using the
relation $\mu$$\approx$(1+$z$)$^{-4}$ where $\mu$ is the galaxy surface
brightness in the SDSS r band via the 50\% \cite{Petrosian76} radii
(petroRad50\_r):
$\mu$=petroMag\_r + 2.5(0.798 + 2log(petroRad50\_r)) and the galaxy $r$--$i$
colors to pick the red galaxies this method is intended for.
The Petrosian radii may add useful information because of the angular
diameter distance relation. We also find this to be the case for GPs as
discussed in Section~\ref{sec:section6} below.

\cite{Carliles08} have used Random Forests (ensembles of classification and
regression trees) to estimate photometric redshifts from the SDSS. Like
GPs (see Paper I) this method is also supposed to give realistic individual
galaxy photometric redshift error estimates and few or no
catastrophic photometric redshift prediction failures.
\cite{Ball08} continue their work using machine learning methods to
derive photometric redshifts for galaxies and quasars using the SDSS and
the Galaxy Evolution
Explorer \citep[GALEX,][]{Martin05}\footnote{http://www.galex.caltech.edu/}.
In particular, they have made interesting progress in eliminating catastrophic
failures in quasar photo-$z$ estimation while bringing down the
rms error (RMSE) values. Work by \cite{kaczmarcik2009} uses astrometric
information to break degeneracies in quasar photometric redshifts which may
also be applied to other kinds of data.

\cite{WG08} have taken a Bayesian approach using the SDSS apparent magnitude
colors $u$--$g$, $g$--$r$, $r$--$i$, $i$--$z$, surface brightness $\mu_{i}$
in the $i$ band,
the S\'{e}rsic $n$--index \citep{Sersic68}, and the absolute magnitude $M_{i}$
``corrected" to z=0.1. Some of these quantities are only available from the
New York University Value Added Catalog (NYC-VAGC) of \citet{Blanton05} or
calculated from the raw photometry directly.
\cite{Wang08} have used support vector machines \citep[also see][]{Wadadekar05}
and kernel regression on a SDSS and Two Micron All Sky Survey
\citep[2MASS,][]{Skrutskie06}\footnote{http://www.ipac.caltech.edu/2mass/}
cross-match list.

\cite{DAbrusco07} utilized a supervised neural network using a standard
multilayer perceptron, but operated in a Bayesian framework on two
different SDSS datasets. One of their data sets consists of
the SDSS Data Release Five \citep[DR5][]{AM07} luminous red galaxy (LRG) sample
\citep{Eisenstein01}, and the other which they term the ``General Galaxy
sample" includes all objects classified as ``GALAXY" in the SDSS.
They then break their sample up into two redshift ranges and after some
interpolation fit to the residuals they obtain impressive results, especially
for the LRG sample (see their Table 4).
In a higher redshift study \cite{Stabenau08} used surface brightness priors
to improve their template based scheme for photometric redshifts in the
VVDS \citep{LeFevre04}\footnote{VLT Very Deep Survey}
and GOODS \citep{Giavalisco04}\footnote{Great Observatories Origins Deep Survey}
surveys.

This certainly does not cover all of the recent work in this field, but is a
representative sample to show the intense interest being generated because
of near-future large-area multi-band surveys like the Large Synoptic
Survey Telescope \citep[LSST][]{Ivezic08}\footnote{http://www.lsst.org}
and PanStarrs \citep{Kaiser02}.

We have used a variety of datasets in our analysis which are discussed
in Section \ref{sec:section2}. Discussion of the photometric and spectroscopic 
quality of the datasets along with other photometric pipeline output 
properties of interest are found in Section \ref{sec:section3}. The methods 
used to obtain photometric redshifts are in Section \ref{sec:section4}. How to 
pick the optimal sample size, matrix rank, and inversion method in 
Section  \ref{sec:section5}. Results are in Section \ref{sec:section6} and 
Conclusions in Section \ref{sec:section7}.

%%% Section 2
\section{The Sloan Digital Sky Survey, The Two Micron All Sky Survey and
The Galaxy Evolution Explorer Datasets}\label{sec:section2}

Most of the work herein utilizes the SDSS Main Galaxy Sample
\citep[MGS,][]{Strauss02} and the LRG
sample \citep[LRG][]{Eisenstein01} from the SDSS Data Release Three
\citep[DR3,][]{Abazajian05} and DR5 \citep{AM07}.
We include the DR3 to facilitate comparison between the present work 
and that from Paper I.  We also utilize the DR5 to maximize the size of
our cross-match catalogs.

For comparison with other work we have cross-matched the SDSS datasets 
with both the 2MASS extended source catalog and GALEX Data Release 4
(GR4)\footnote{http://galex.stsci.edu/GR4} All Sky Survey photometric 
attributes.  Our method of cross-matching these catalogs has not changed 
since Paper I except that we now cross-match against the SDSS DR5 instead
of the DR3 to increase the size of our catalogs. Many aspects of the SDSS,
2MASS, and GALEX surveys relevant to this work were described in Paper I
and hence we will not repeat them here. The only new catalog included since
Paper I is the SDSS LRG. The SDSS LRG sample
is similar to the SDSS MGS except that it explicitly targets the LRGs.
These galaxies have a fairly uniform spectral energy distribution
(SED) and a strong 4000 {\AA} break which tend to make calculating
photometric redshifts easier than for the MGS \citep[e.g.][]{Padmanabhan05}
since the training--set contains more homogenous SEDs.
Since these galaxies are among the most luminous
galaxies in the universe and tend to be found in over dense regions
(e.g., clusters/groups of galaxies) they are also good candidates
for mapping the largest scales in the universe; see \citep{Eisenstein01}
for more details.

\section{Photometric and redshift quality, morphological indicators and
other catalog properties}\label{sec:section3}

For SDSS photometric and redshift quality we follow much the same recipe as
in Paper I. However, unlike Paper I we refrain from using SDSS photometry
of the highest quality (what we referred to as ``GREAT") as we did not see
any consistent improvments in our regression fits using this higher quality
photometry. We stick with the SDSS photometric ``GOOD" flags as defined in
Paper I: !BRIGHT and !BLENDED and !SATURATED.  See Table 2 in Paper I for a
description of the flags.  We utilize the same photometric quality flags
for the GALEX and 2MASS datasets as described in Paper I, Section 3.
We incorporate the same SDSS morphological indicators as in our previous work
(See Paper I, Section 3.5). The SDSS casjobs\footnote{http://casjobs.sdss.org}
queries used to get the data are the same as those in the Appendix of Paper I
except in the case of the LRGs utilized herein which require
primtarget=TARGET\_GALAXY\_RED (p.primtarget \& 0x00000020 $>$ 0) instead
of primtarget=TARGET\_GALAXY (p.primtarget \& 0x00000040 $>$ 0) for the MGS.

Tables~\ref{tbl-1} and \ref{tbl-2} contain a comprehensive list of the six data
sets used herein.
%\begin{deluxetable}{l|l|l}
\begin{deluxetable}{lll}
\tablecolumns{3}
%\rotate
\tablewidth{0pc}
\tablecaption{Data Sets 1-3\label{tbl-1}}
\tabletypesize{\scriptsize}
\tablehead{
\colhead{Data Set 1\tablenotemark{a}} &
\colhead{Data Set 2} &
\colhead{Data Set 3} \\
\colhead{SDSS-DR3 MGS} &
\colhead{SDSS-DR5 LRG} &
\colhead{SDSS-DR3 MGS + GALEX-GR4} \\
\colhead{Training=180045,Testing=20229\tablenotemark{b}} &
\colhead{Training=87002,Testing=9666} &
\colhead{Training=30036,Testing=3374}
}
\startdata
$g$-$r$-$i$                         & $g$-$r$-$i$                       & $g$-$r$-$i$\\
$u$-$g$-$r$-$i$                     & $u$-$g$-$r$-$i$                   & $u$-$g$-$r$-$i$\\
$g$-$r$-$i$-$z$                     & $g$-$r$-$i$-$z$                   & $g$-$r$-$i$-$z$\\
$u$-$g$-$r$-$i$-$z$                 & $u$-$g$-$r$-$i$-$z$               & $u$-$g$-$r$-$i$-$z$\\
...                                 & ...                               & nuv-fuv-$g$-$r$-$i$\\
...                                 & ...                               & nuv-fuv-$u$-$g$-$r$-$i$\\
...                                 & ...                               & nuv-fuv-$g$-$r$-$i$-$z$\\
...                                 & ...                               & nuv-fuv-$u$-$g$-$r$-$i$-$z$\\
$u$-$g$-$r$-$i$-$z$-p50             & $u$-$g$-$r$-$i$-$z$-p50           & nuv-fuv-$u$-$g$-$r$-$i$-$z$-p50\\
$u$-$g$-$r$-$i$-$z$-p50-p90         & $u$-$g$-$r$-$i$-$z$-p50-p90       & nuv-fuv-$u$-$g$-$r$-$i$-$z$-p50-p90\\
$u$-$g$-$r$-$i$-$z$-p50-p90-ci      & $u$-$g$-$r$-$i$-$z$-p50-p90-ci    & nuv-fuv-$u$-$g$-$r$-$i$-$z$-p50-p90-ci\\
$u$-$g$-$r$-$i$-$z$-p50-p90-ci-qr   & $u$-$g$-$r$-$i$-$z$-p50-p90-ci-qr & nuv-fuv-$u$-$g$-$r$-$i$-$z$-p50-p90-ci-qr\\
$u$-$g$-$r$-$i$-$z$-p50-p90-fd      & $u$-$g$-$r$-$i$-$z$-p50-p90-fd    & nuv-fuv-$u$-$g$-$r$-$i$-$z$-p50-p90-fd\\
$u$-$g$-$r$-$i$-$z$-z-p50-p90-fd-qr & $u$-$g$-$r$-$i$-$z$-p50-p90-fd-qr & nuv-fuv-$u$-$g$-$r$-$i$-$z$-p50-p90-fd-qr\\
\enddata
\tablenotetext{a}{$u$-$g$-$r$-$i$-$z$=5 SDSS magnitudes, p50=Petrosian 50\% light radius
in the SDSS $r$ band, p90=Petrosian 90\% light radius in the $r$ band,
ci=Petrosian inverse concentration index, fd=FracDev value, qr=Stokes $Q$ value
in the $r$ band, nuv=GALEX Near UV band, fuv=GALEX Far UV band,
see Paper I Section 3.6 for more details.}
\tablenotetext{b}{These are the sizes of the testing and training sets used
in our analysis}
\end{deluxetable}

\begin{deluxetable}{lll}
\tablecolumns{3}
%\rotate
\tablewidth{0pc}
\tablecaption{Data Sets 4-6\label{tbl-2}}
\tabletypesize{\scriptsize}
\tablehead{
\colhead{Data Set 4\tablenotemark{a}} &
\colhead{Data Set 5} &
\colhead{Data Set 6} \\
\colhead{SDSS-DR5 LRG + GALEX-GR4} &
\colhead{SDSS-DR5 MGS + 2MASS} &
\colhead{SDSS-DR5 LRG + 2MASS} \\
\colhead{Training=4042,Testing=454} &
\colhead{Training=133947,Testing=15050} &
\colhead{Training=39344,Testing=4420} 
}
\startdata
$g$-$r$-$i$                               & $g$-$r$-$i$                 & $g$-$r$-$i$\\
$u$-$g$-$r$-$i$                           & $u$-$g$-$r$-$i$             & $u$-$g$-$r$-$i$\\
$g$-$r$-$i$-$z$                           & $g$-$r$-$i$-$z$             & $g$-$r$-$i$-$z$\\
$u$-$g$-$r$-$i$-$z$                       & $u$-$g$-$r$-$i$-$z$         & $u$-$g$-$r$-$i$-$z$\\
nuv-fuv-$g$-$r$-$i$                       & $g$-$r$-$i$-$j$-$h$-$k$     & $g$-$r$-$i$-$j$-$h$-$k$\\
nuv-fuv-$u$-$g$-$r$-$i$                   & $u$-$g$-$r$-$i$-$j$-$h$-$k$ & $u$-$g$-$r$-$i$-$j$-$h$-$k$\\
nuv-fuv-$g$-$r$-$i$-$z$                   & $g$-$r$-$i$-$z$-$j$-$h$-$k$ & $g$-$r$-$i$-$z$-$j$-$h$-$k$\\
nuv-fuv-$u$-$g$-$r$-$i$-$z$               & $u$-$g$-$r$-$i$-$z$-j-h-k   & $u$-$g$-$r$-$i$-$z$-$j$-$h$-$k$\\
nuv-fuv-$u$-$g$-$r$-$i$-$z$-p50           & ...                         & ...\\
nuv-fuv-$u$-$g$-$r$-$i$-$z$-p50-p90       & ...                         & ...\\
nuv-fuv-$u$-$g$-$r$-$i$-$z$-p50-p90-ci    & ...                         & ...\\
nuv-fuv-$u$-$g$-$r$-$i$-$z$-p50-p90-ci-qr & ...                         & ...\\
nuv-fuv-$u$-$g$-$r$-$i$-$z$-p50-p90-fd    & ...                         & ...\\
nuv-fuv-$u$-$g$-$r$-$i$-$z$-p50-p90-fd-qr & ...                         & ...\\
\enddata
\tablenotetext{a}{$u$-$g$-$r$-$i$-$z$=5 SDSS magnitudes, p50=Petrosian 50\% light radius
in the SDSS $r$ band, p90=Petrosian 90\% light radius in the $r$ band, ci=Petrosian
inverse concentration index, fd=FracDev value, qr=Stokes $Q$ value in $r$ band,
nuv=GALEX Near UV band,
fuv=GALEX Far UV band, $j$=2MASS $j$ band, $h$=2MASS $h$ band, $k$=2MASS $k$ band;
see Paper I Section 3.6 for more details.}
\end{deluxetable}

\section{Improved Gaussian Process Methods}\label{sec:section4}

In this section we will discuss our investigation of different GP
transfer functions (kernels) \& rank-reduction matrix inversion 
techniques.  Our results suggest that there may be an upper limit to the
number of training-set galaxies needed to derive photometric redshifts
using the SDSS, but this result should be viewed with caution.
While there have been recent suggestions that one may quantify the maximum
number of galaxies required to obtain an optimal fit \citep{BH2009},
in practice what we see with the GPs could be an artifact of the algorithm
itself. In particular, it might be desirable to explore building good 
``local" models to compare with the present GPs (and neural networks),
which are global models.

In the GP method utilized herein one would begin with a 
training set matrix $X$ of dimensions $n \times d$, where $n$ is the number 
of galaxies and $d$ is the number of components which might include broad 
band flux measurements and morphological information.  
One would also have a target vector $y$ of dimensions $n \times 1$, which
would contain the known redshift for each galaxy in our case. The testing 
data are in a matrix $X^{*}$ of dimension $n^{*} \times d$
with target values in a matrix $y^{*}$ consisting of 
$n^{*} \times 1$ redshifts, where $n^{*}$ is the number of test samples.
We wish to predict the value of
$y^{*}$ given as $X$, $y$, and $X^{*}$. The prediction of $y^{*}$ requires a
covariance function $k(x,x')$, with $x$ and $x'$ vectors with $d$ components.
This covariance function can be used to construct a $n \times n$ covariance 
matrix $K$,
where $K_{ij}=k(x_{i},x_{j})$ for rows $x_{i}$ and $x_{j}$ of $X$, and the 
$n^{*} \times n$ cross covariance
matrix $K^{*}$ ($K^{*}_{ij}=k(x^{*}_{i},x_{j})$ where $x^{*}_{i}$ is the
$i$th row of $X^{*}$). Once this is accomplished, the prediction 
$\hat{y}^{*}$ for $y^{*}$ may be given by the GP equation 
\citep[][p. 17]{RW2006}:
\begin{equation}
\hat{y}^{*}=K^{*}(\lambda^{2}I + K)^{-1}y \label{eq:GP}
\end{equation}
where $\lambda$ represents the noise in $y$ and can be used to improve the 
quality of the model \citep{RW2006}. 

In addition to the prediction $\hat{y}^{*}$, the GP approach
also leads to an equation for $C$ the covariance matrix for the predictions in
equation \ref{eq:GP}. If the $n^{*} \times n^{*}$ matrix $K^{**}$ has
entries $K^{*}_{ij}=k(x^{*}_{i},x^{*}_{j})$ then \citep[][p. 79]{RW2006}:

\begin{equation}
C=K^{**}-K^{*}(\lambda I + K)^{-1}K^{*T}
\end{equation}

The superscript $T$ indicates the transpose. The pointwise variance of
the prediction is diag($C$), the diagonal of the  $n^{*} \times n^{*}$ matrix
$C$.

For details about the selection of 
$\lambda$, the covariance function (kernel) $k$, hyperparameters in the 
kernel, and GPR in general see 
\cite{foster09} and \cite{RW2006}.  The following discussion is a summary of 
\cite{foster09}.  We will use the above notation for the sections that 
follow.

\subsection{Different Kernel choices}

In Paper I we relied exclusively on a polynomial kernel, but to investigate
the possibility that other kernels might perform better we have tried several
other common forms in the meantime.

The squared exponential (SE) kernel function (also known as the 'radial
basis' kernel function) is given by
\begin{equation}
k_{SE}(r) = \exp\left( - \frac{r^2}{2 l^2}\right)
\end{equation}
where $l$ is the length scale. The length scale determines the rate at which 
the kernel function drops to zero away from the origin.  This covariance 
function is infinitely differentiable and hence is very smooth. Because it
is so smooth, it can sometimes be unrealistic for use in modeling real 
physical processes. 

The Matern class covariance function is given by
\begin{equation}
k(r) = \frac{2^{l-v}}{\Gamma(v)}\left(\frac{\sqrt{2vr}}{l}\right)^v K_{v}\left(\frac{\sqrt{2vr}}{l}\right)
\end{equation}
where $v$ and $l$ are positive parameters and $K_{v}$ is a modified Bessel
Function.  As $v \rightarrow \infty$ this reduces to the SE
above.  The process becomes very non-smooth for $v = \frac{1}{2}$ and for
values of $v \geq \frac{7}{2}$, the function is as rough as noise. The Matern
class covariance function is mean square differentiable $k$ times if and only
if $v > k$.  The Matern class of covariance functions can be used
to model real physical processes and is more realistic than the above SE
covariance function.

The rational quadratic covariance function is given by
\begin{equation}
k(r) = \left(1 + \frac{r^{2}}{2 \alpha l^{2}}\right)^{-\alpha}
\end{equation}
As the value of the parameter $\alpha \rightarrow \infty$ this reduces to 
the SE function described earlier.  Unlike the Matern class covariance
function, this function is mean square differentiable for every value of
$\alpha$.

The polynomial covariance function is given by
\begin{equation}
k(x,x') =(\sigma_{0}^{2} + x^{T}\Sigma_{p}x')^{p}
\end{equation}
where $\Sigma_{p}$ is a positive semidefinite matrix and $p$ is a positive
integer.  If $\sigma_{0}^{2} = 0$ the kernel is homogeneous and linear,
otherwise it is inhomogeneous. In principle this function may not be suitable
for regression problems as the variance grows with 
$\mid x \mid$ for $\mid x \mid > 1$. 
However there
are applications where it has turned out to be effective \citep{RW2006}.

The neural network covariance function is given by
\begin{equation}
k_{NN}(x,x') = \frac{2}{\pi} sin^{-1}
\left(\frac{2x^{T}\Sigma x'}
{\sqrt{(1+2x^{T}\Sigma{x})(1+2x'^{T}\Sigma x')}}\right) \label{eq:NN}
\end{equation}
This covariance function is named after neural networks because the function
can be derived from the limiting case of a model of a neural
network \citep{Neal1996}

In our calculations we chose $\Sigma$, which scales as the training-set data,
to have the form $I/l^{2}$ where $I$ is a $d \times d$ identity matrix.
The hyperparameters $l$ and $\lambda$ were selected by finding a (local)
maximum to the marginal likelihood using the routine $minimize$ from 
\citet[][pp. 112-116, 221]{RW2006}.

Two or more covariance functions can be combined to produce a new covariance
function. For example sums, products, convolutions, tensor products and
other combinations of covariance functions can be used to form new covariance
functions. Details are described in \cite{RW2006}.

For the calculations shown in the rest of the paper we utilized 
equation \ref{eq:NN}, the neural network kernel, since for our data
it outperformed all other kernels.

\subsection{Low Rank Approximation Matrix Inversion
Techniques}\label{sec:subsection4.2}

As mentioned in Paper I (Section 4.4) to utilize GPR
the inversion of the matrix $M=(\lambda^{2}I + K)$ in equation \ref{eq:GP}
is required.  This matrix turns out to be an $n\times n$ non-sparse matrix where
$n$ is the number of training-set galaxies.  Paper I mentioned that matrix
inversion requires $O(n^{3})$ floating point operations. Thus,
to accommodate the matrix in memory and to keep the computation feasible,
we kept $n\leq$1000 in Paper I.

This was a severe shortcoming for GPs since they had 1--2 orders of magnitude
less training samples to work with than all of the other methods described in
Paper I.  Nonetheless, GPs performed extremely well within this limitation.

Since writing Paper I, we have developed a variety of rank-reduction methods
to invert large non-sparse matrices. These will make GPR much more competitive
than that shown in Paper I. \cite{foster09} outline the rank-reduction methods
utilized in detail, so we provide a brief summary of their advantages below.

Note that the number of samples, $n$, is the same as that described above, while
the rank, $m<n$, is the size of the rank-reduced matrix.  We typically keep 
$m<$1500 to keep the numbers of operations to invert the matrices manageable 
in wall-clock time. Memory usage for the methods below is $O(nm)$.

\noindent SR-N: the subset of regressors method.
This method has been proposed and utilized in the past
\citep{RW2006,wahba1990,PG1990} and requires nm$^{2}$ flops
to invert. However, this method is known to have problems with numerical
stability. That problem is addressed in the methods below.

\noindent SR-Q: the subset of regressors using a QR factorization.
The use of the QR factorization \citep[][p.239]{golub1996} is designed
to reduce computer arithmetic errors in the SR-N method.
This method requires 2nm$^{2}$ flops to invert.
Therefore, it is a little more expensive than SR-N.

\noindent SR-V: the $V$ method. 
Since this method in combination with pivoting (see below)
is the one we utilize the most in later aspects
of this paper we will go into a little more depth here.
From Section \ref{sec:section4} Equation (\ref{eq:GP}) we recall
that the size of $(\lambda^{2}I + K)^{-1}$ is $n \times n$ and as
mentioned above for large $n$ it is not practical to calculate
$(\lambda^{2}I + K)^{-1}$ directly.  To get around this we will
approximate $K$ with $VV^{T}$ where $V$ is produced by
partial Cholesky factorization (see \cite{foster09}).
Let $K^{*}_{1}$ be the first $m$ columns of $K^{*}$ and let $V_{11}$
be the $m \times m$ matrix of the first $m$ rows of $V$ where $m < n$.
Then let $V^{*} = K^{*}_{1}V^{-T}_{11}$. In addition to replacing
$K$ with $VV^{T}$ we can also approximate $K^{*}$ with $V^{*}V^{T}$.
With these substitutions one sees that  $K^{*}(\lambda^{2}I + K)^{-1}y$
from Equation (\ref{eq:GP}) can be approximated by
$V^{*}V^{T}(\lambda^{2}I + VV^{T})^{-1}y$. It turns out that this can also
be written as $\hat{y}^{*}=V^{*}(\lambda^{2}I + V^{T}V)^{-1}V^{T}y$.
The matrix $(\lambda^{2}I + V^{T}V)^{-1}$ is now $m \times m$ instead
of $n \times n$ and for small enough $m$ the equation can be solved
quite quickly. The new flop count will be $O$(nm$^{2}$).

This method is intermediate in terms of growth of computer arithmetic
errors between the normal equations and the SR-Q method, but in general
the accuracy is close the SR-Q. This method was first discussed
by \cite{seeger2003} and \citet[][p.136]{wahba1990}.

\noindent SR-NP, SR-QP, SR-VP: the use of pivoting with rank-reduction methods.
All of the previous methods use the first m columns of $K$, but one can
select any subset of the columns to construct a low--rank approximation.
Selecting these columns is part of the problem to be solved. Our approach is
similar to that of \cite{fine2001}.

Pivoting is useful in forming a numerically stable low--rank approximation of
a positive semi-definite matrix, and to do so it identifies the rows of the
training data which limit the growth of computer arithmetic errors. A pivot of
the matrix $K$, which is simply a permutation of $K$ of the form $PKP^{T}$
corresponds to the permutation $PX$ of $X$. It is possible to move columns
and rows of $K$ so that the $m \times m$ leading principal submatrix
of $PKP^{T}$ has the condition number that is a function of $n$ and $m$.
Thus pivoting will tend to construct a low--rank approximation whose condition
number is related to the condition number of the low--rank approximation
produced by the singular--value decomposition. However, the growth
of computer arithmetic errors in the algorithm depends on the condition number
of the low--rank approximation. Since pivoting limits the condition number
and the growth of computer arithmetic errors depends on the condition number,
pivoting will tend to improve the numerical stability of the algorithm. This
can, in principle, reduce the effect of computer arithmetic errors. If computer
arithmetic errors are larger than the other errors (such as measurement errors
and modelling errors) in the prediction of the redshift, then an algorithm
incorporating pivoting may potentially be more accurate than an algorithm
without pivoting.

Examples 2--4 in \cite{foster09} illustrate some of the
dangers of not pivoting and how they are resolved with pivoting for small
(artificial) problems.

In the end adding pivoting increases
SR-N to 2nm$^{2}$ flops and SR-Q to 3nm$^{2}$ while SR-V stays the same.

\section{Comparison: Picking the optimal Sample Size, Rank size, and
Matrix Inversion Method}\label{sec:section5}

Here we investigate Data Set 1 in detail in order to discern a variety
of things including: is there an optimal sample size for a given survey;
what is the best matrix inversion method; if using rank-reduction methods
what is the optimal rank size? When discussing conventional matrix
inversion, we will be limited to a maximum of 20,000 training samples
\footnote{This is due to memory(RAM) limitations. Our 64-bit compute
platform is based around a 2 x 2.66 Ghz Dual-Core Intel Xeon with 16GB of
667Mhz DDR2 RAM}. 

\begin{figure}[Figure1]
\includegraphics{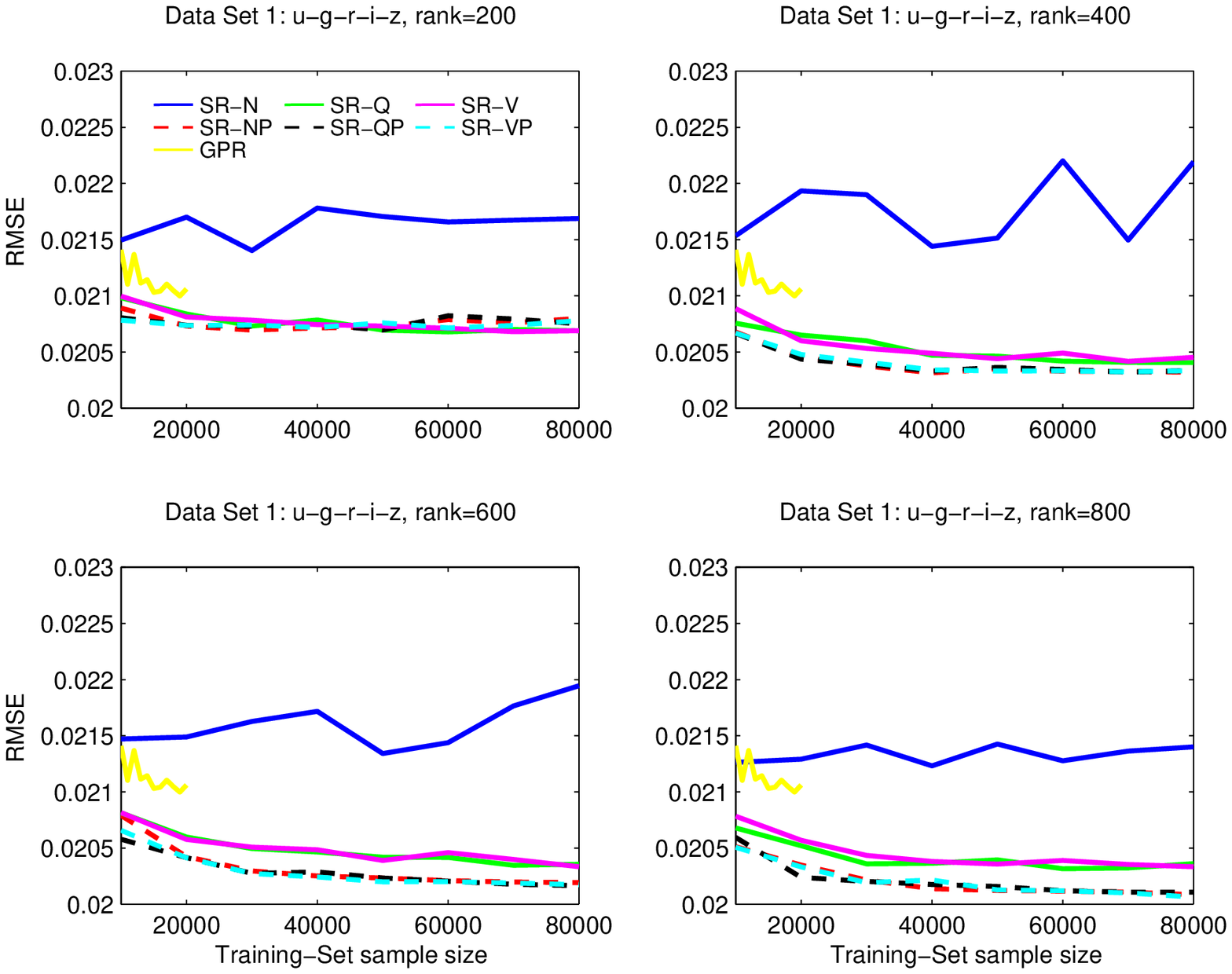}
\caption{From Data Set 1 (see Table~\ref{tbl-1}). Error bars are not plotted
for reasons of clarity; however, they are of the same
order as the scatter in the lines.}\label{fig:Figure1}
\end{figure}

\begin{figure}[Figure2]
\centering
\includegraphics{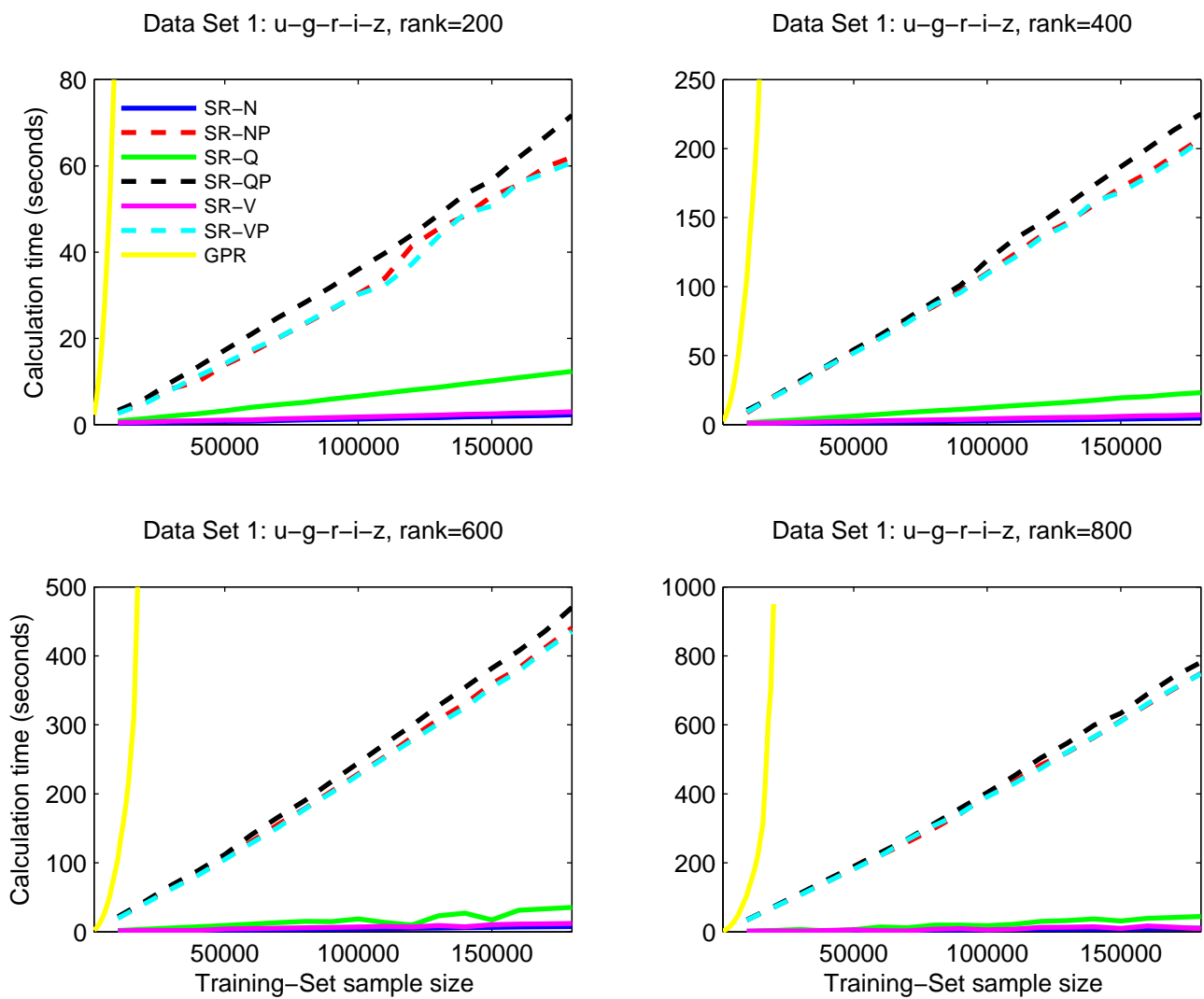}
\caption{From Data Set 1 (see Table~\ref{tbl-1}), but unlike in Figure 1
we show that the matrix inversion times are linear out to the full size
(180,000 galaxies) of the data set.}\label{fig:Figure2}
\end{figure}

Figures \ref{fig:Figure1} and \ref{fig:Figure2} show the variation of
RMSE and calculation time versus
sample size.  For the GP method (which is labeled GPR 
and is in yellow), this involved a full matrix inversion up to 20,000 
training-set samples. The rest of the curves are from the other
rank-reduction matrix inversion techniques and are labeled as described in the 
previous section.  Several features are apparent:

\begin{enumerate}

\item The SR-N method does not perform well in comparison to any of the
other techniques. However, it does invert its matrices much faster than
the standard matrix inversion technique.

\item Except for the SR-N method, all of the other rank-reduction methods
outperform the full matrix reduction in the range of 10,000--20,000 samples.

\item The rank-reduction methods with pivoting slightly outperform the
non-pivoting methods in term of lower RMSE values. However, the pivoting
methods take much more time to do the matrix inversions than the
non-pivoting methods.

\item More training-set samples give lower RMSE values. By around 40,000
samples the curves start to level off regardless of the rank size.

\item Larger rank sizes clearly give better performance in terms
of lower RMSE for a given sample size.  This is described in more detail 
below.

\end{enumerate}

\begin{figure}[Figure3]
\centering
\includegraphics{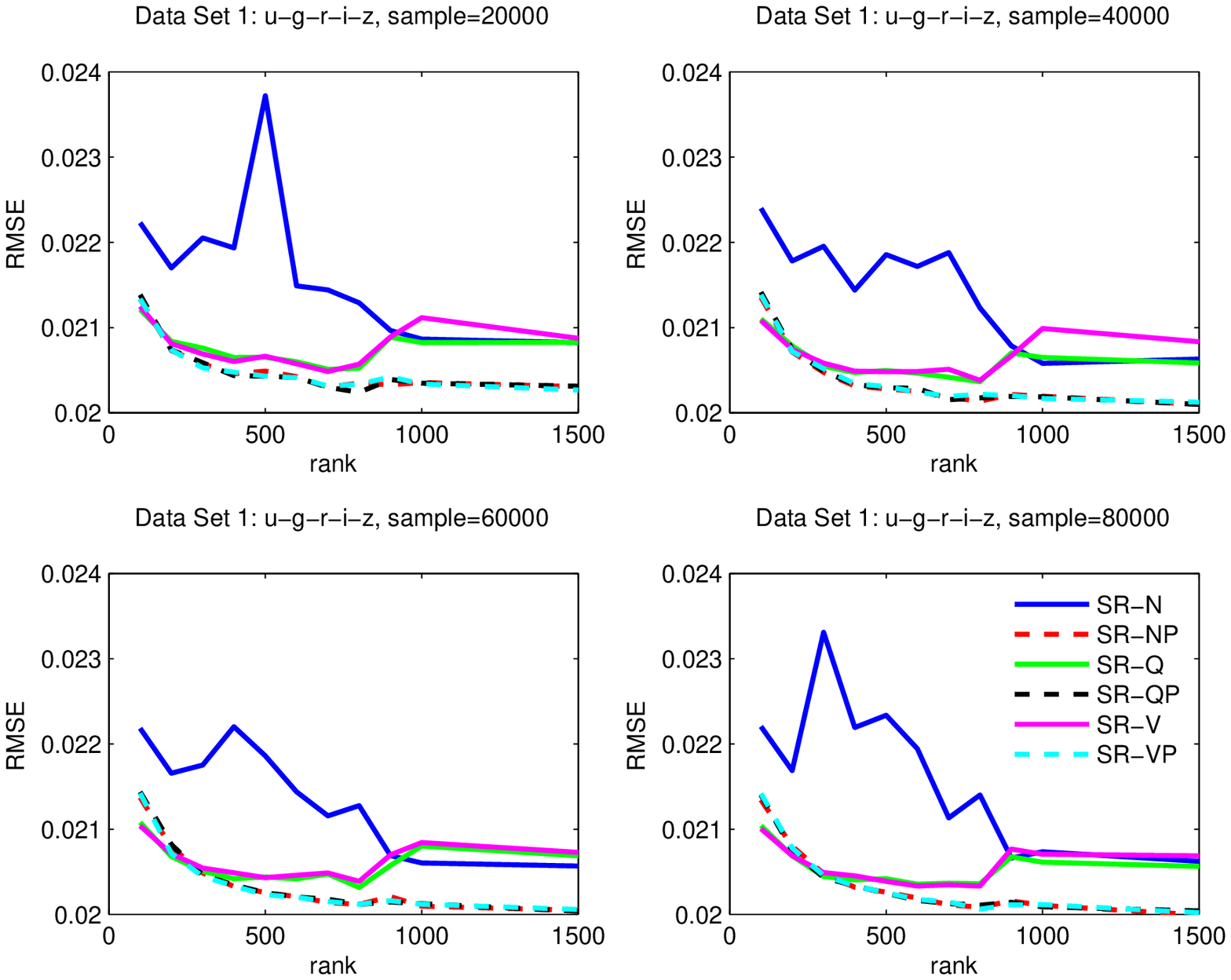}
\caption{From Data Set 1 (see Table~\ref{tbl-1}) error bars are not plotted
for reasons of clarity. They are of the same
order as the scatter in the lines.}\label{fig:Figure3}
\end{figure}

Figure \ref{fig:Figure3} shows the variation of RMSE with rank for 
several different sample sizes.  The rank is plotted from 100 to 1000 in 
increments of 100, but we also add rank=1500 to see if there is a large 
change in calculated RMSE for a much larger value. Some important 
features to note here:

\begin{enumerate}

\item As in Figure \ref{fig:Figure1}, the RMSE decreases for larger
sample sizes, but as was noted earlier, there is not a large difference
between sample sizes of 40,000 and above.

\item For the non-pivoting matrix inversion techniques (not
including SR-N) SR-Q and SR-V the RMSE increases beyond rank=800. This
suggests that there might be some instability associated with non-pivoting
methods as rank size becomes large.  For this reason, one should stick with 
the pivoting methods (SR-QP or SR-VP) if one wishes to use a rank of 800 or 
larger.

\item On average it appears that SR-VP and SR-QP outperform the other
rank reduction methods. SR-VP also appears to outperform SR-QP, although
the difference is marginal.

\item SR-VP with rank=800 and sample size=40000 appear to be optimal
choices for our data when looking at Figures
\ref{fig:Figure1}--\ref{fig:Figure3} given the accuracy of the result.
The timings are much longer for these pivoting methods as shown above,
but they outperform all other methods.

\end{enumerate}

\section{Results}\label{sec:section6}

\subsection{SDSS Main Galaxy and LRG Results}

\begin{figure}[Figure4]
\plottwo{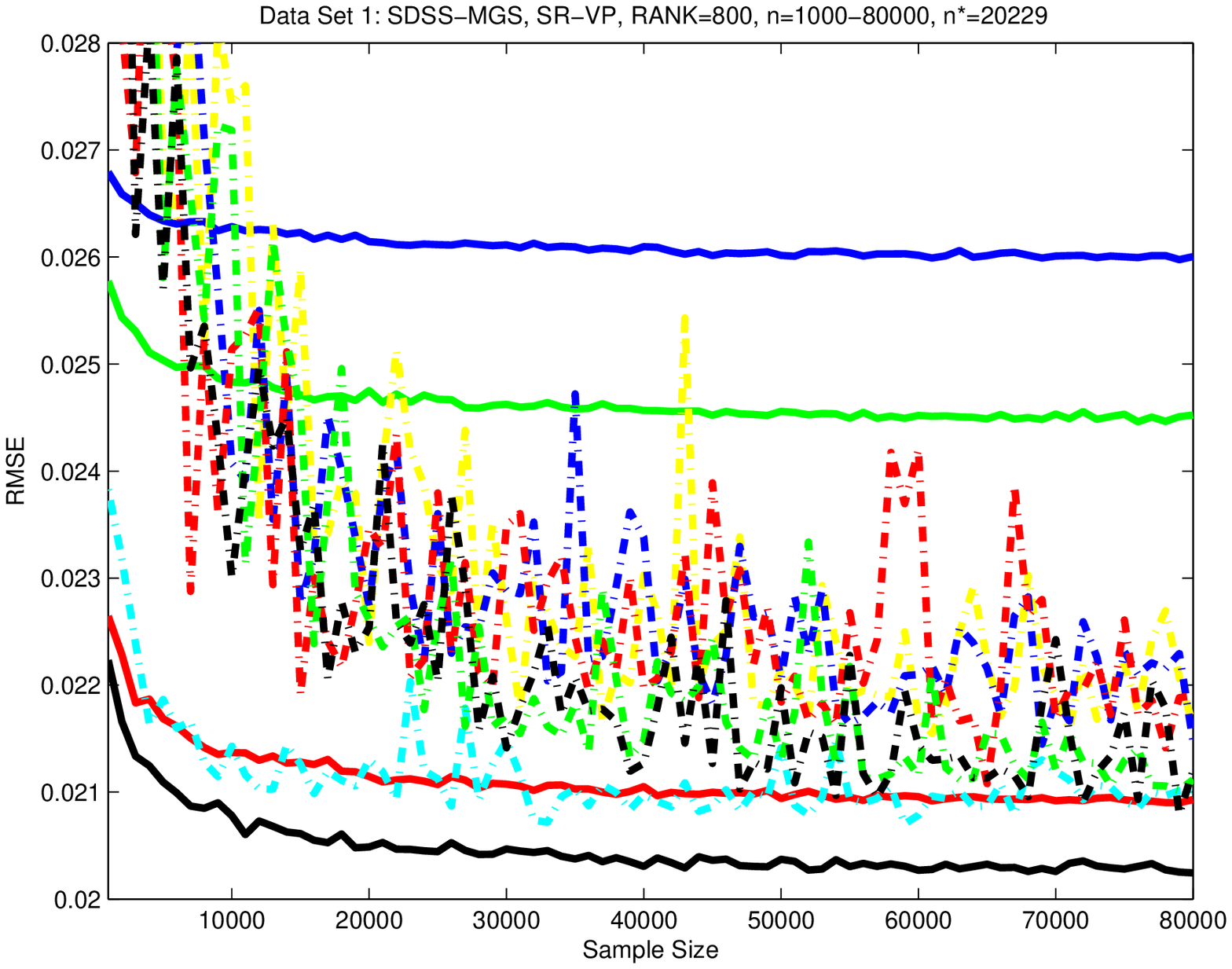}{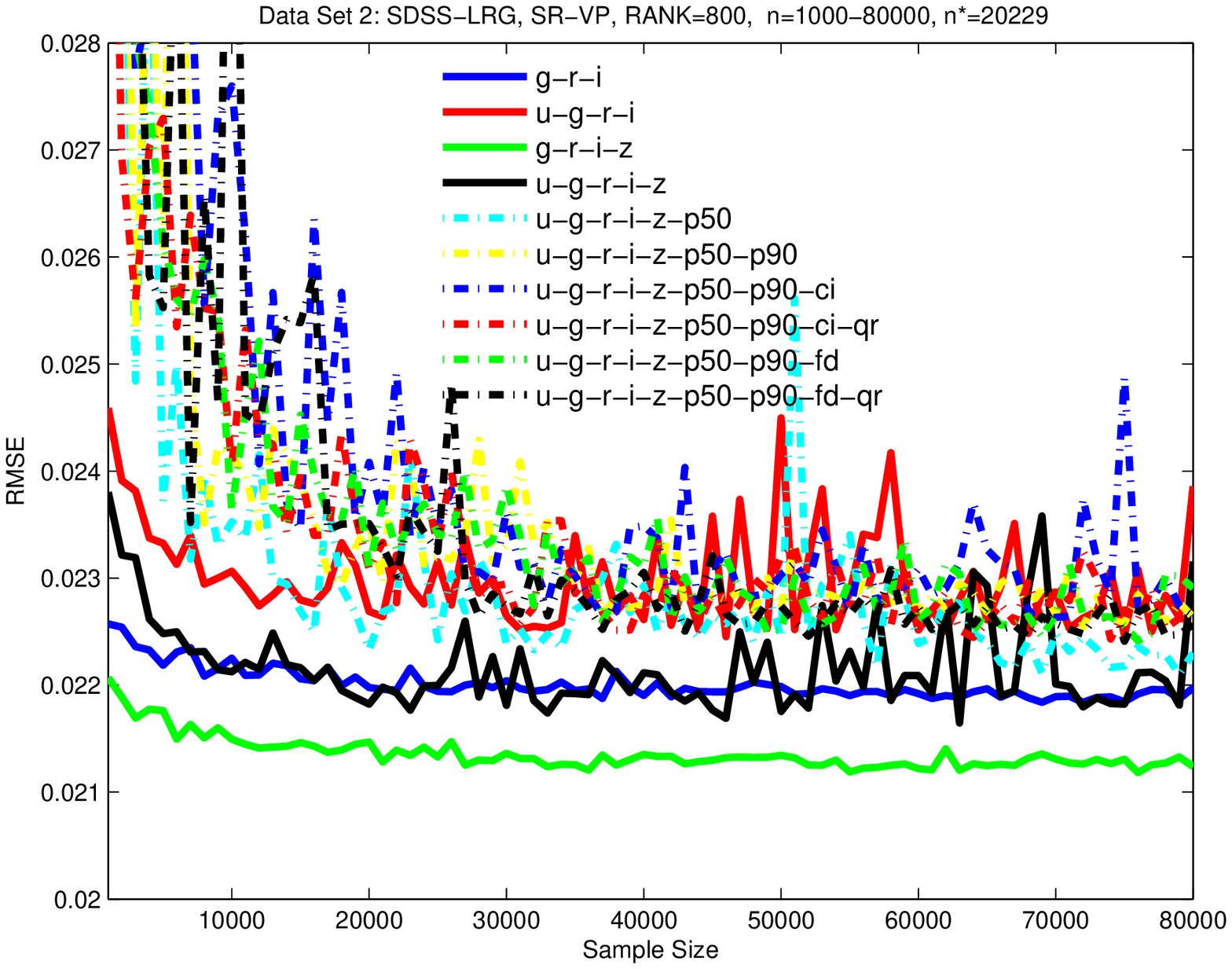}
\caption{\scriptsize{From Data Sets 1 and 2 (see Table~\ref{tbl-1}). We utilize
the rank--reduction method termed SR-VP with a rank size of 800. The training
sets ($n$ in the plot, following our earlier notation) range in size from 1000
to 80,000 in 1000 increments with 10 bootstraps \citep{ET93} per run.
The testing sample size ($n^{*}$) was always 20,229. The mean value of
the 10 bootstraps is plotted. 90\% confidence levels from the bootstrap
resampling are of order the vertical line variation. Clearly, the errors
are much larger for those which include the morphological parameters.}
}\label{fig:Figure4}
\end{figure}

The SDSS MGS (Data Set 1) \& LRG (Data Set 2) will give us different
results because the LRG sample has far fewer SED types than are
found in the SDSS MGS while the LRG sample goes to fainter magnitudes and
hence deeper redshifts (see Figures \ref{fig:Figure8} and \ref{fig:Figure9}).
This will make the job of any regression algorithm 
quite different. This is evident in the two panels of Figure \ref{fig:Figure4},
which show the variation of RMSE versus sample size for the two different data
sets.  A number of points need to be stressed:

\begin{enumerate}
\item $Morphological Inputs$. The morphological information
(p50, p90, ci, fd, qr) may add some information that the regression algorithm
can utilize.  This includes the Petrosian 50\% radii (p50), the Petrosian 90\%
(p90), the inverse concentration index (ci=p50/p90), the FracDev (fd) and
Stokes $Q$ parameter (qr) all in the SDSS $r$ band. More details on these
parameters are discussed in Paper I. Data Set 1
(Figure \ref{fig:Figure4}(a)) and the five SDSS filters
$u$-$g$-$r$-$i$-$z$ (not including morphology inputs)
clearly outperform all of the subsets of $u$-$g$-$r$-$i$-$z$ ($g$-$r$-$i$,
$u$-$g$-$r$-$i$, and $g$-$r$-$i$-$z$) and the addition of morphological
inputs. In Data Set 2 (Figure \ref{fig:Figure4}(b)) the morphological
information appears to add noise for the most part making the fits worse than
by using only combinations of the five SDSS $u$-$g$-$r$-$i$-$z$ bandpass
filters.

\item $Fewer SEDs$. As mentioned in the previous section, by the time sample
sizes of $\sim$ 40,000 are reached in the SDSS--MGS of Data Set 1
(Figure \ref{fig:Figure4}(a)) the RMSE begins to level off. In the SDSS--LRG of
Data Set 2 (Figure \ref{fig:Figure4}(b)) however this is already occurring
for most of the inputs in the 10,000--20,000 range.
This is clearly the advantage of
having less SEDs to worry about in the SDSS--LRG sample versus the SDSS--MGS.
In fact for Data Set 2 (SDSS--LRG) it is clear that only four of the five SDSS
bandpasses are sufficient for the optimal fit ($g$-$r$-$i$-$z$). The SDSS $u$
bandpass is clearly superfluous in the SDSS--LRG data set when using
GP fitting routines.

\item $Errors$. 90\% confidence levels derived from the bootstrap resampling
are roughly at the level of the variation in each of the inputs used as a
function of sample size. It is clear that adding morphological information
requires larger error estimates for these datasets.
\end{enumerate}

\subsection{Cross-Matching GALEX and SDSS Results}

\begin{figure}[Figure5]
\plottwo{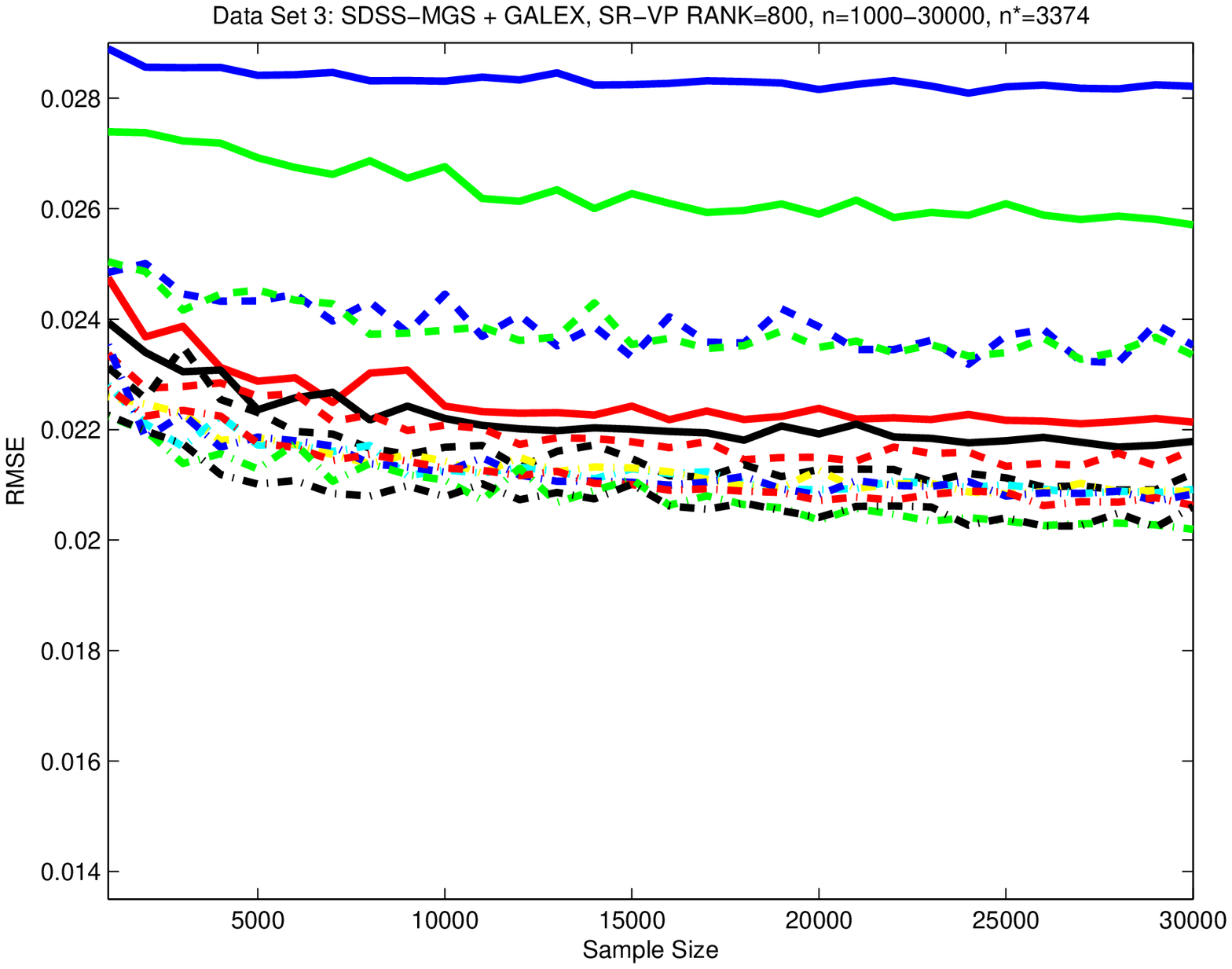}{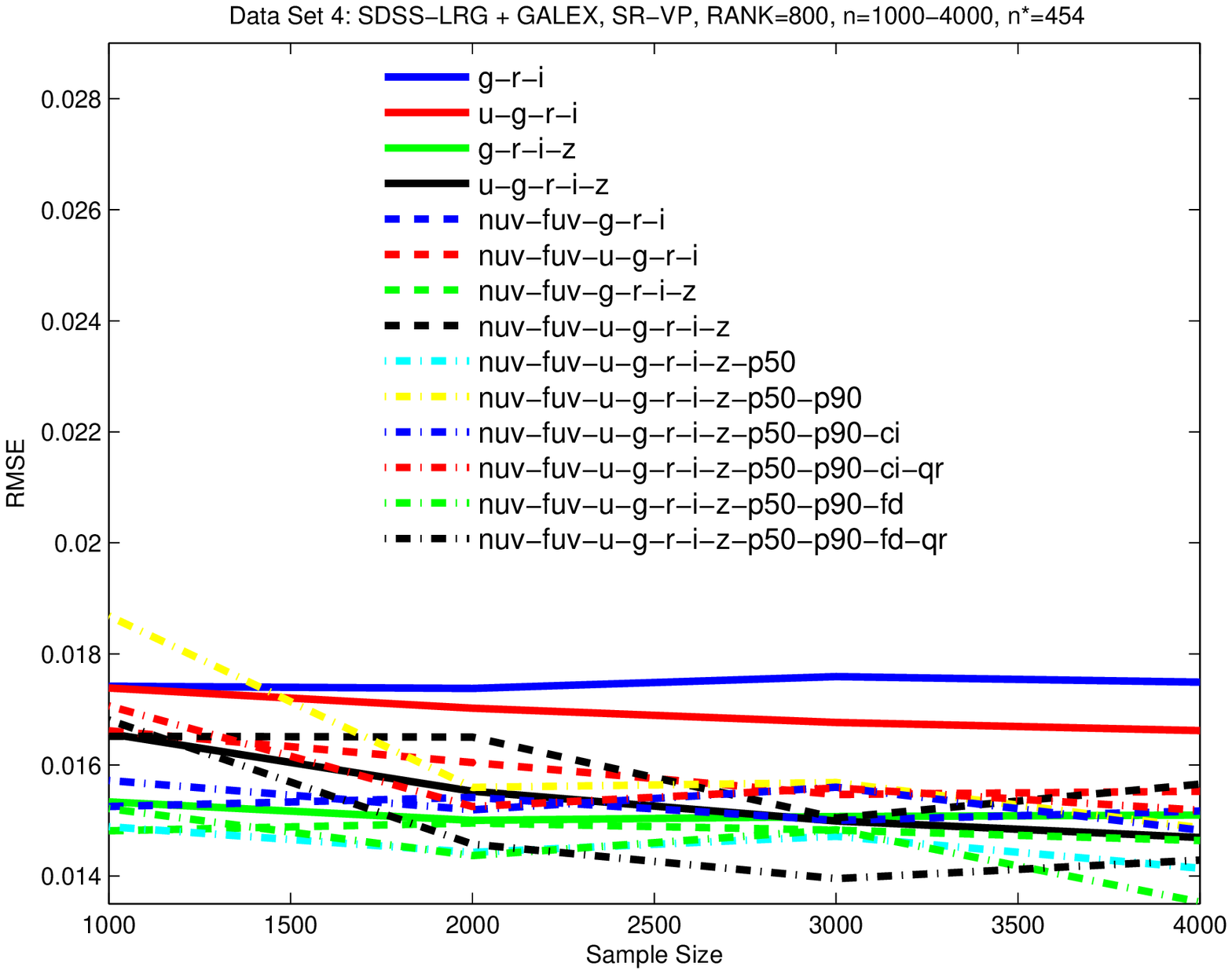}
\caption{\scriptsize{From Data Sets 3 and 4 (see Table~\ref{tbl-1}). We utilize
the rank-reduction method termed SR-VP with a rank size of 800. On the left in
plot (a), we use training sets ($n$ in the plot, following our earlier notation)
ranging in size from 1000 to 30,000 in 1000 increments with 10 bootstraps
per run. The testing sample size ($n^{*}$) is 3374. The mean value of 10
bootstraps resampling runs is plotted. 90\% confidence levels from the bootstrap
resampling are of order the vertical line variation.
On the right, we use similar notation, but we have smaller training
(1000--4000 in increments of 1000)
and testing (454) sets.}}\label{fig:Figure5}
\end{figure}

\begin{figure}[Figure6]
\plottwo{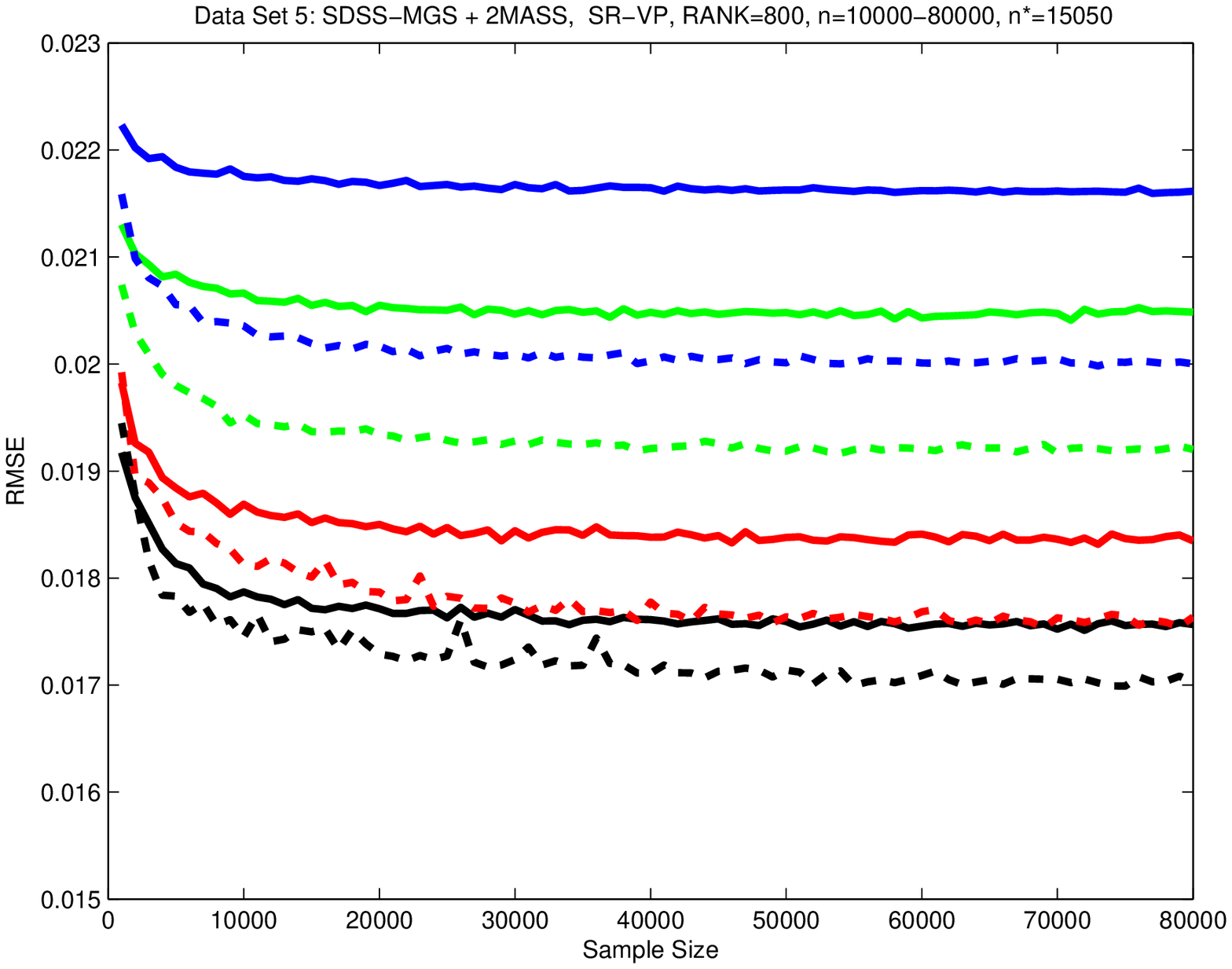}{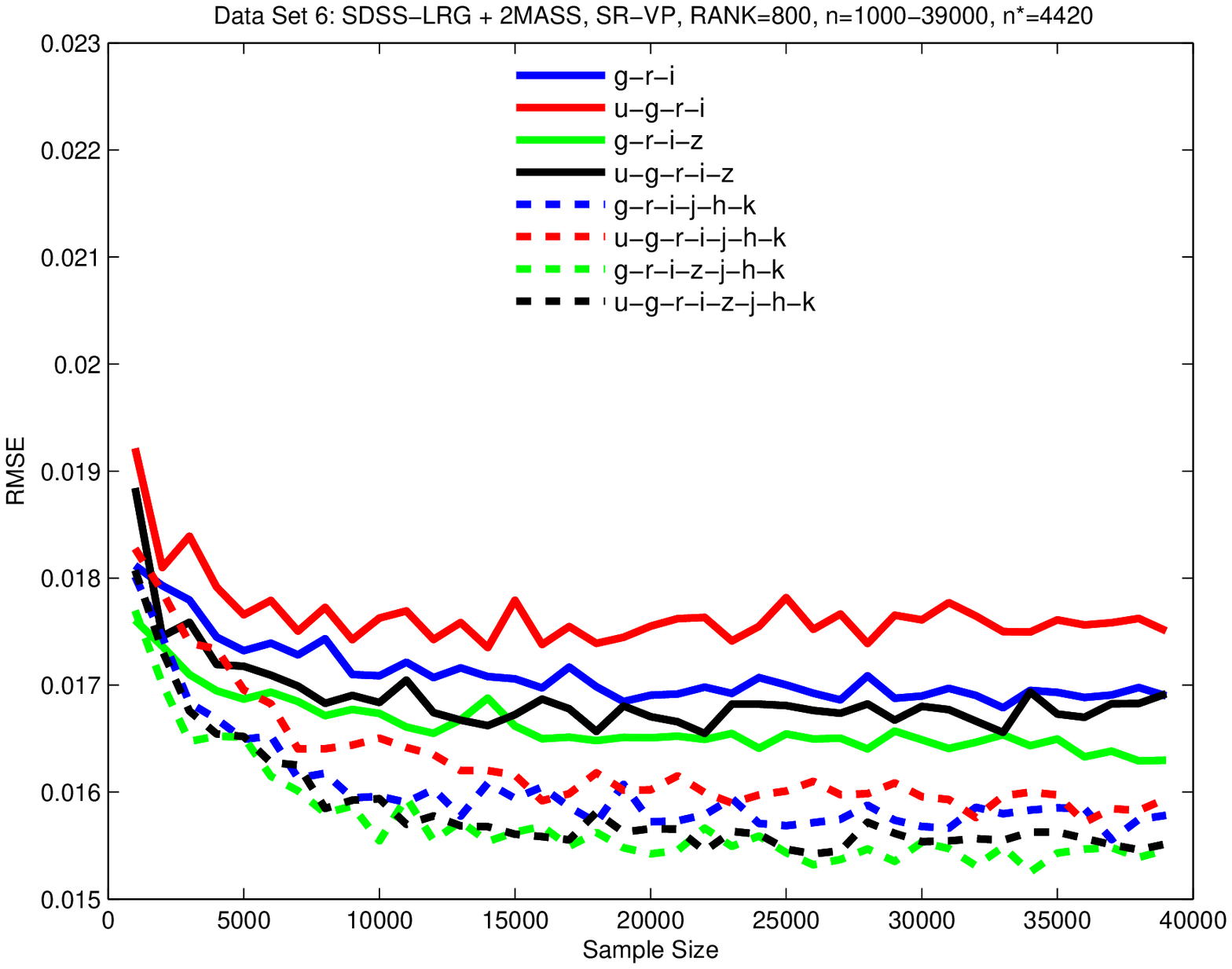}
\caption{From Data Sets 5 and 6 (see Table~\ref{tbl-1}). We utilize
the rank--reduction method termed SR-VP with a rank size of 800. For Data Set 5
the training sets (denoted as $n$) range in size from 1000 to 80,000 in 1000
increments with 10 bootstraps per run and a testing-set ($n^{*}$) size of
15,050.
On the right, Data Set 6 training sets range from 1000 to 40,000 in increments
of 1000 with 10 bootstraps per run and a testing-set size of 4420. Bootstrap
90\% confidence levels are again of order the vertical line variation.}
\label{fig:Figure6}
\end{figure}

Figure \ref{fig:Figure5} shows results from a cross-match of the SDSS
and GALEX catalogs, which are listed as Data Sets 3 and 4 in Tables~\ref{tbl-1}
and \ref{tbl-2}. Figure \ref{fig:Figure7} shows the SDSS and SDSS + GALEX
results for Data Sets 1--4, but without any SDSS morphological inputs
included.  This is to better quantify the differences between the SDSS and
SDSS + GALEX GP fits.  The following should be noted:

\begin{enumerate}

\item Comparing Figure \ref{fig:Figure4}(a) to Figure \ref{fig:Figure5}(a)
one sees that those inputs that include SDSS morphological information
are slightly improved when GALEX filters are included. The
error bars on those with morphological inputs (errors not shown here) are also
smaller in Figure \ref{fig:Figure5}(a) versus Figure \ref{fig:Figure4}(a). This
would imply that the addition of GALEX filters helps make better use of the
morphological inputs.

\item Figure \ref{fig:Figure7}(a) is made up of Figures \ref{fig:Figure4}(a),
\ref{fig:Figure5}(a), and \ref{fig:Figure6}(a) without the SDSS morphological
information included.  One notices that Data Set 3 (SDSS--MGS + GALEX)
in Figure \ref{fig:Figure7}(a) has higher RMSE values for the purely SDSS
bandpasses ($g$-$r$-$i$, $u$-$g$-$r$-$i$, $g$-$r$-$i$-$z$,
$u$-$g$-$r$-$i$-$z$) than Data Set 1 (SDSS--MGS only).  Here the max size of the
training data sets is different by a factor of 2.7 (80,000 versus 30,000)
hence the difference may be attributed to a smaller data set size, although
that is unlikely given how we subsample the data in Data Set 1. However, if one
examines Figure \ref{fig:Figure8} one sees clear differences and similarities
in the magnitude and redshift distributions of these two catalogs. In particular
the r-band magnitude distribution is quite distinct, the z-band less so.
This seems to have made it harder for the GPs to obtain a good fit
for the MGS galaxies. Within Data Set 3 of Figure \ref{fig:Figure7}(a) the
GALEX bandpasses help with two of the SDSS only input options ($g$-$r$-$i$ and
$g$-$r$-$i$-$z$) compared to Data Set 1. However, the two GALEX bandpasses
do not help with the best inputs from Data Set 1 ($u$-$g$-$r$-$i$
and $u$-$g$-$r$-$i$-$z$). Hence for the MGS
galaxies there appears no need to utilize the GALEX magnitudes to improve
photo-z estimation over that already obtained from SDSS only magnitudes.
The same applies to the the SDSS morphological information, which adds very
little of substance. For example, compare u-g-r-i-z in Data Set 1
(Figure \ref{fig:Figure4}(a)) versus
nuv-fuv-$u$-$g$-$r$-$i$-$z$-p50-p90-fd-qr in Data Set 3
(Figure \ref{fig:Figure5}(a)).

\item Comparing Figures \ref{fig:Figure4}(b) and \ref{fig:Figure5}(b),
one sees that
the LRG + GALEX cross-match catalog has lower RMSE values than the LRG
only catalog regardless of the inputs used. Hence one would be led to believe
that one should always use GALEX magnitudes where available for LRG galaxies
to improve photo-$z$ estimation. However, there are two other things to take
note of. First, one again sees that the max training data set size is a factor
of 20 smaller (80,000 versus 4000) between Data Sets 4 and 2, although Data Set
2 does take a subsample at the level of Data Set 4. Therefore, sample size does
not appear to be the issue here.  Looking at Figure \ref{fig:Figure9}
it is clear that there are few similarities in the magnitude or redshift
distributions for these two data sets. Clearly the GP algorithm is fitting
a completely different set of data points and it finds Data Set 4 much
easier than Data Set 2.

\item Looking at Figure \ref{fig:Figure7}(b) (made up of Figures
\ref{fig:Figure4}(b), \ref{fig:Figure5}(b) and \ref{fig:Figure6}(b)
without the SDSS
morphological inputs included) the addition of the GALEX nuv--fuv filters
within Data Set 4 seem to assist in photo-$z$ estimation when using SDSS filters
$g$-$r$-$i$ and $u$-$g$-$r$-$i$, but has a little effect when added to the
already superior $g$-$r$-$i$-$z$ and $u$-$g$-$r$-$i$-$z$.

\end{enumerate}

As noted above, the RMSE differences between Figures \ref{fig:Figure4}(a) and
\ref{fig:Figure5}(a) suggest that the underlying distribution of SDSS magnitudes
and redshifts of Data Set 1 versus 3 are different as seen
in Figure \ref{fig:Figure8}.  The data set has shrunk in size between
Data Sets 1 and 3,
while the redshift distribution appears the same. However, the colors
of the galaxies have changed enough that the GPs find it harder with the
reduced sample size to obtain a good fit.

The explanation for the improvement seen between Figures \ref{fig:Figure4}(b)
and \ref{fig:Figure5}(b) (Data Sets 2 and 4) is perhaps simpler.
Figure \ref{fig:Figure9} shows the $u$,$r$,$z$ and redshift distributions for
these two data sets. Clearly, the centroid, spread,
and shape of the distributions of the $u$,$r$,$z$ and redshift distributions
are signficantly different. The LRG + GALEX redshift distribution in
particular is strongly truncated beyond a redshift of about 0.2
while the magnitude distributions tend to be more Gaussian in shape.
Certainly it is easier for GPs to come up with better fits
for lower-redshift distributions. 

The marked differences between the SDSS MGS and LRG results are because of the
different galaxy SEDs that exist in each catalog. These differences also exist
because the LRG samples go fainter than the MGS samples
\citep[see][]{Eisenstein01} and they have a different redshift and galaxy
magnitude distribution (see Figures \ref{fig:Figure8} and \ref{fig:Figure9}).
The magnitude and redshift differences between the pure LRG
and LRG+GALEX catalogs are much larger than they are between
the corresponding MGS and MGS+GALEX catalogs.  Clearly the additional
GALEX inputs affect the SDSS MGS only ($u$-$g$-$r$-$i$-$z$)
results negatively, while the GALEX inputs affect on the LRG sample is
ambiguous at best. These differences suggest that one must be very careful
in interpreting the improvement in RMSE results associated with any
SDSS + GALEX cross-match catalogs.

\subsection{Cross-Matching 2MASS and SDSS Results}

Figure \ref{fig:Figure6} demonstrates our GPR results from a cross-match
catalog containing the 2MASS extended source catalog with the SDSS MGS
(Data Set 5) and the SDSS LRG sample (Data Set 6). When 
Figure \ref{fig:Figure6} is compared with 
Figure \ref{fig:Figure4}, the results in Figure \ref{fig:Figure6} are 
significantly better for both cases.
While it might be tempting to attribute this improvement to the inclusion
of additional bandpasses in the analysis in Figure \ref{fig:Figure6}, it 
is important to take note of a variety of other important differences 
between the RMSE estimates in these two figures.

\begin{figure}[Figure7]
\plottwo{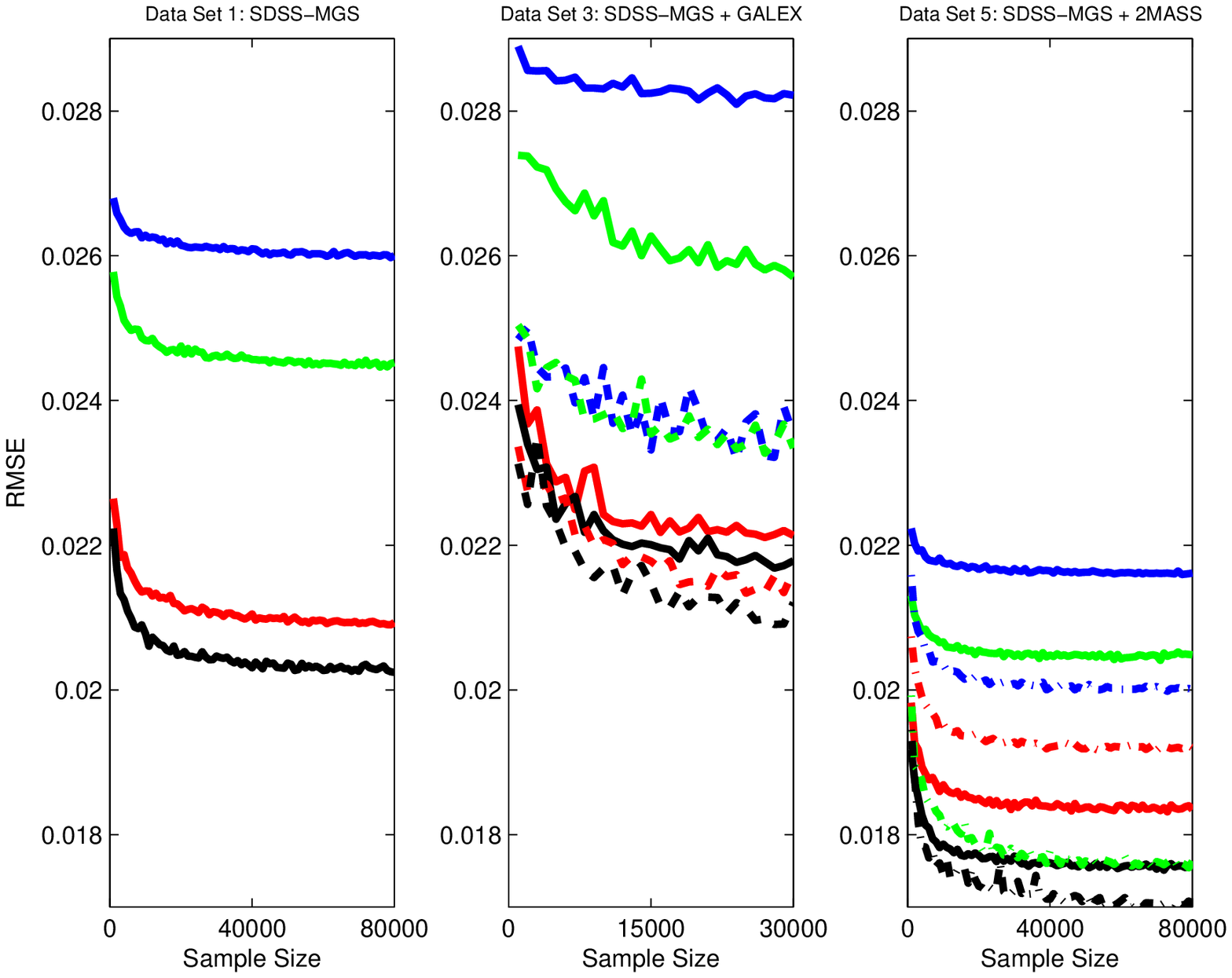}{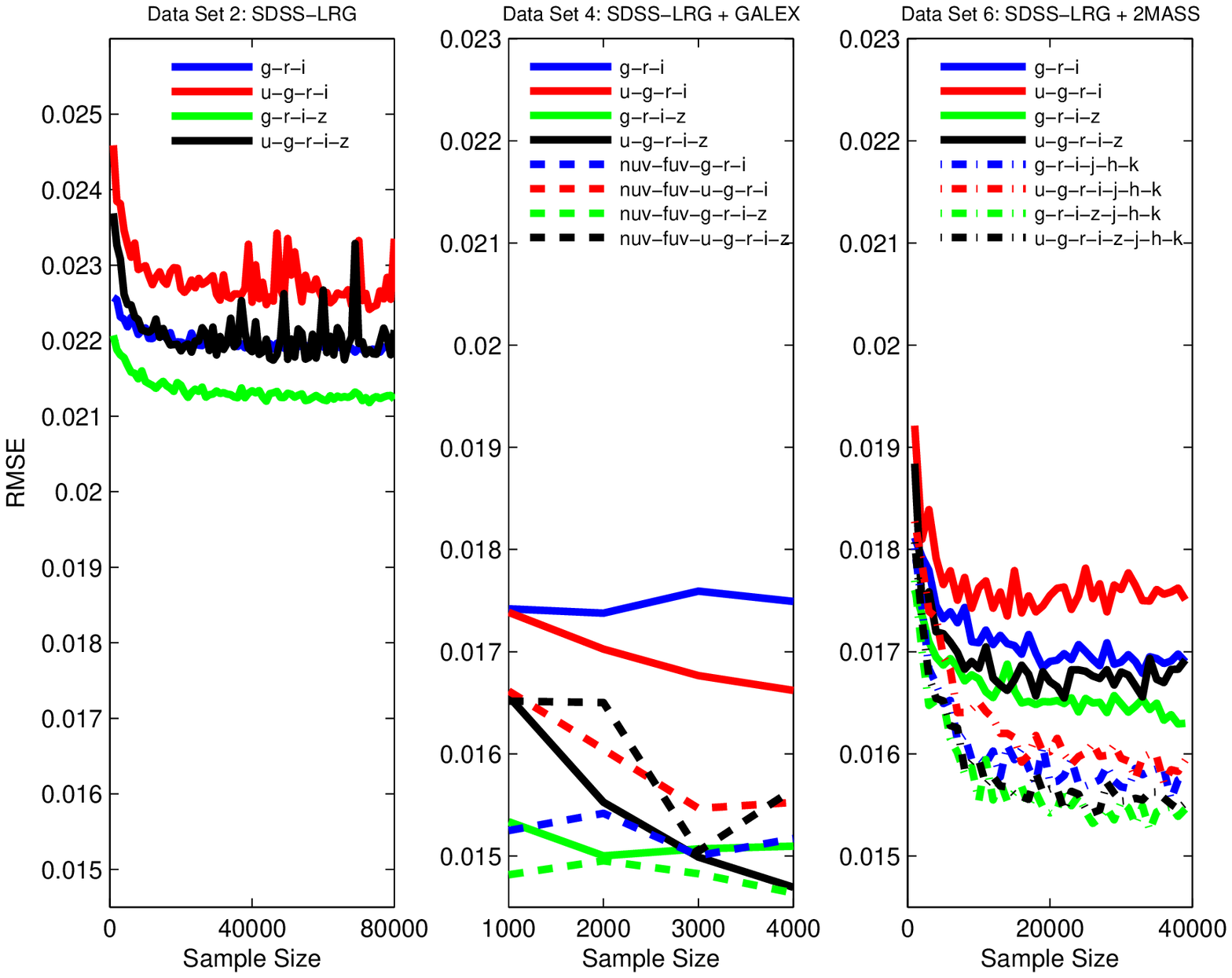}
\caption{From Data Sets 1--6 (see Table~\ref{tbl-1}).
The SDSS $u$-$g$-$r$-$i$-$z$ filter combinations alone along with those
of GALEX nuv, fuv filters, and 2MASS $j$,$h$,$k$. This demonstrates how the
addition
of the GALEX and 2MASS filters influence the SDSS only magnitude fits via
the GP SR-VP method.}\label{fig:Figure7}
\end{figure}

\begin{enumerate}

\item For the SDSS only bandpasses ($u$-$g$-$r$-$i$-$z$) the RMSE drops
significantly
between Data Sets 1--5 (Figure \ref{fig:Figure4}(a) -- \ref{fig:Figure6}(a)) and
Data Sets 2--6 (Figures \ref{fig:Figure4}(b) -- \ref{fig:Figure6}(b));
see Figure \ref{fig:Figure7} for another viewpoint.
This drop is because the 2MASS galaxies
tend to be brighter and at lower redshift making the cross-match catalog
between the 2MASS and SDSS also brigher and lower redshift than
the SDSS only catalog especially for the case of the LRG cross-match
samples (see Figures \ref{fig:Figure10} and \ref{fig:Figure11}).

\item Figure \ref{fig:Figure6}(b) (Data Set 6) has lower RMSE values compared
to Figure \ref{fig:Figure4}(b) (Data Set 2) regardless of input.
It also appears to converge to a best fit RMSE very quickly in comparison
to Data Set 5 (Figure \ref{fig:Figure6}(a)).

\item In Figure \ref{fig:Figure7}(a) (focusing on Data Sets 1 and 5) it is clear
that adding the 2MASS fluxes improves the RMSE fit regardless of which 
SDSS filters are combined with the 2MASS $j$-$h$-$k$ bandpasses.

\item In Figure \ref{fig:Figure6}(b) (Data Set 6) adding 
the 2MASS fluxes can improve the RMSE fit, but the conditions under which
this improvement occurs are significantly different from those in
Figure \ref{fig:Figure6}(a) (Data Set 5).  Upon close inspection
it can be seen that equivalent best results are obtained as the
training sample reaches $\sim$ 20,000 using
$g$-$r$-$i$-$z$-$j$-$h$-$k$ (dashed green).
This shows that for Data Set 6, the $u$ band adds little to the LRG sample.
This is consistent with the behavior observed in Figure \ref{fig:Figure4}(b)
(Data Set 2).

\end{enumerate}

\begin{figure}[Figure8]
\plotone{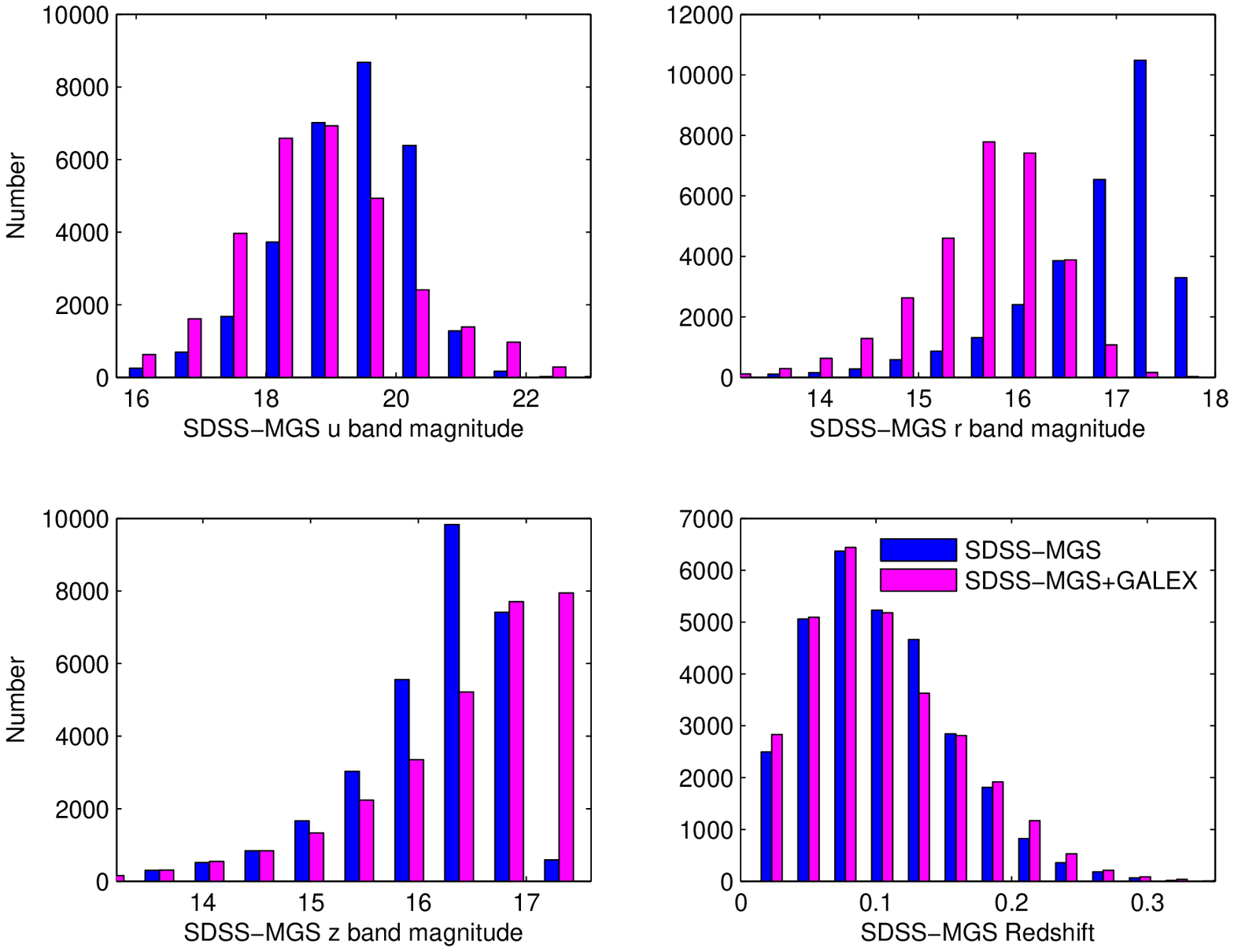}
\caption{Overlapping histograms for Data Sets 1 and 3 (see Table~\ref{tbl-1})
from three of the five SDSS magnitudes ($u$,$r$,$z$).
Data Set 1 is in blue, and Data Set 2
in magenta. Of course, the SDSS+GALEX cross-match catalogs (Data Set 3) are
smaller, so the SDSS only data (Data Set 1) was randomly resampled to be the
same size as the cross-match catalog so that trends in the plots are
directly comparable.}
\label{fig:Figure8}
\end{figure}

\begin{figure}[Figure9]
\plotone{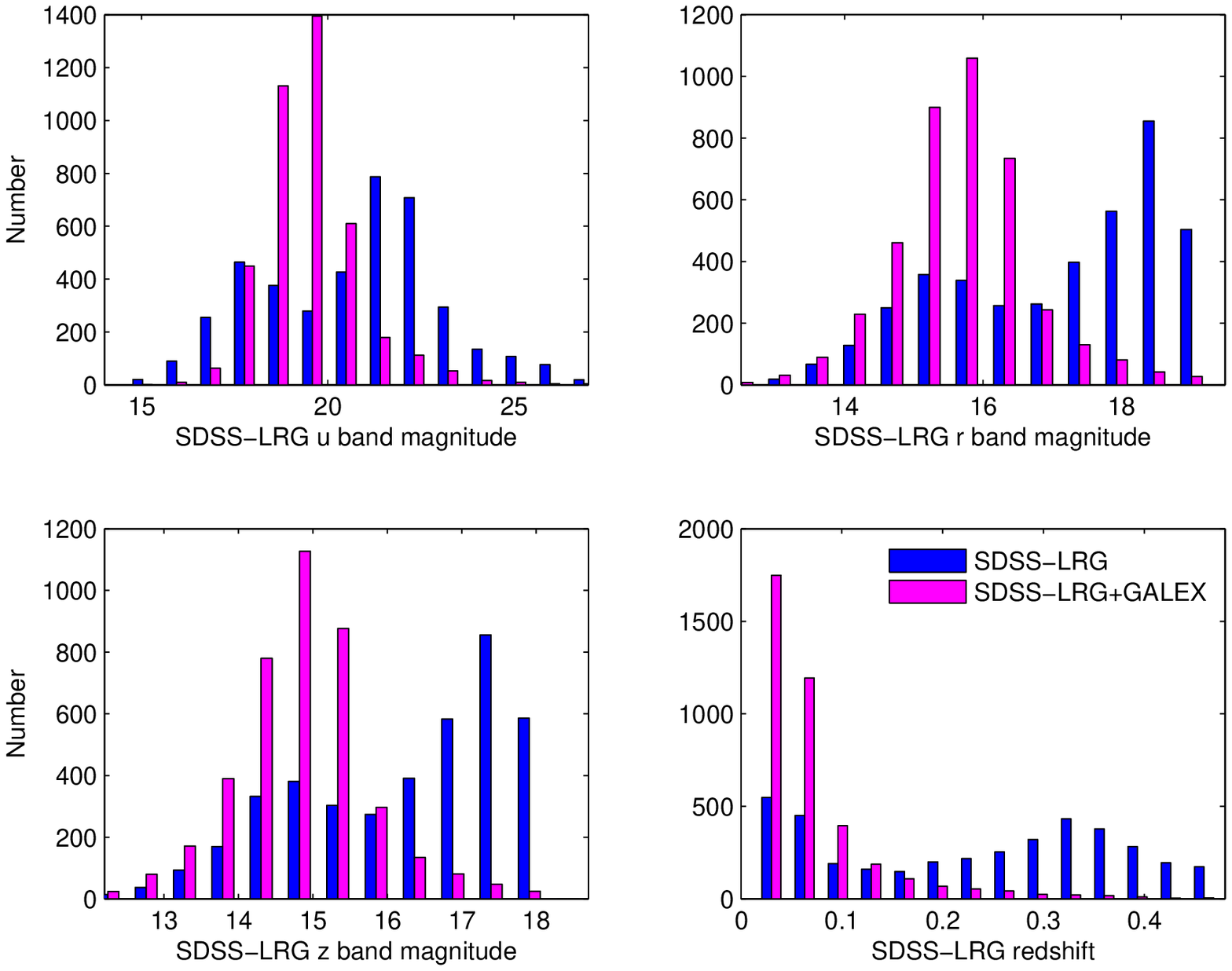}
\caption{Overlapping histograms for Data Sets 2 and 4 (see Table~\ref{tbl-1})
from three of the five SDSS magnitudes ($u$,$r$,$z$).
Data Set 2 is in blue and Data Set 4
in magenta. Of course, the SDSS+GALEX cross-match catalogs (Data Set 4) are
smaller, so the SDSS only data (Data Set 2) was randomly resampled to be the
same size as the cross-match catalog so that trends in the plots are
directly comparable.}
\label{fig:Figure9}
\end{figure}

\begin{figure}[Figure10]
\plotone{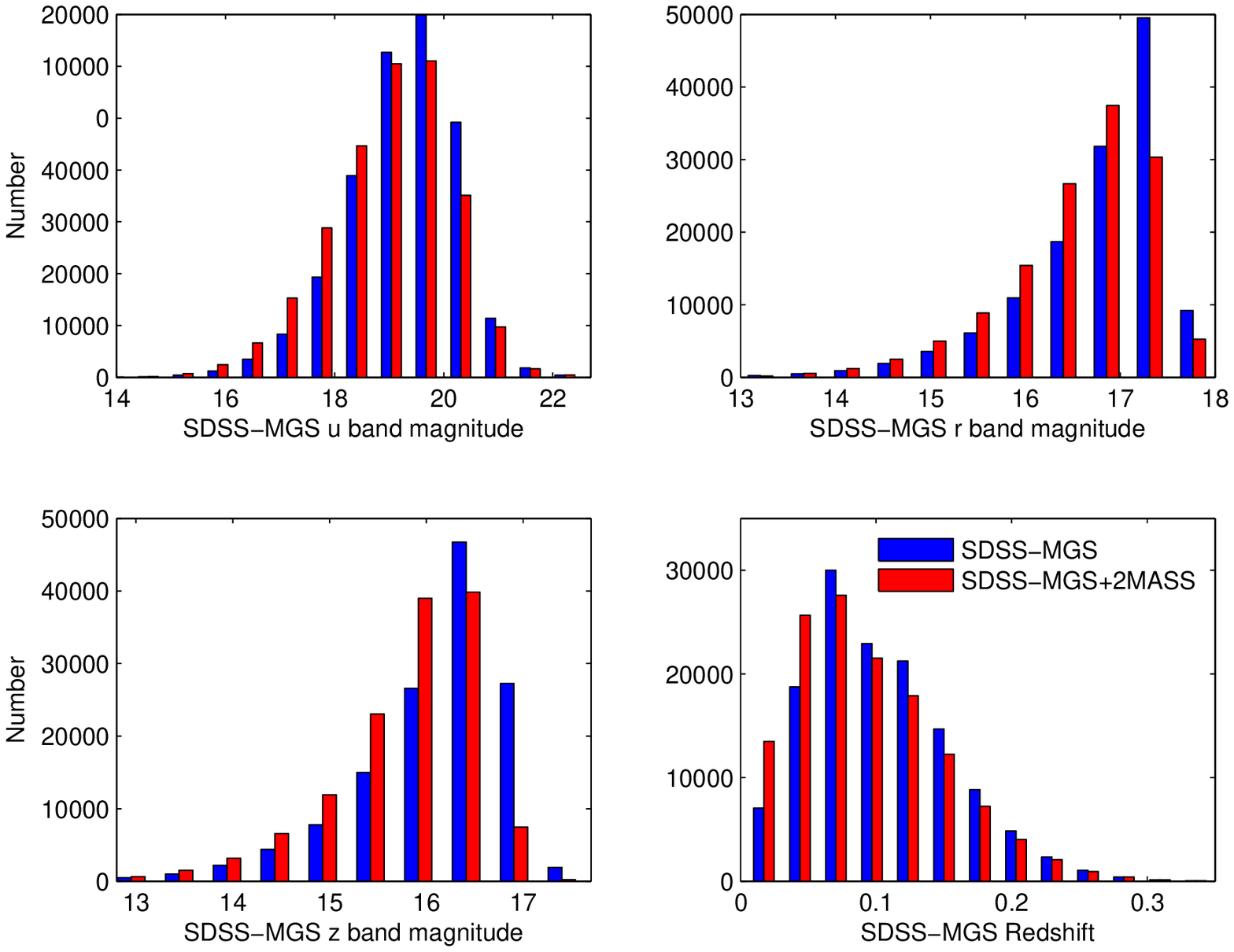}
\caption{Overlapping histograms for Data Sets 1 and 5 (see Table~\ref{tbl-1})
from three of the five SDSS magnitudes ($u$,$r$,$z$). Data Set 1 is in blue and
Data Set 5
in red. Of course, the SDSS+2MASS cross-match catalogs (Data Set 5) are smaller,
so the SDSS only data (Data Set 1) was randomly resampled to be the same size as
the cross-match catalog so that trends in the plots are directly comparable.}
\label{fig:Figure10}
\end{figure}

\begin{figure}[Figure11]
\plotone{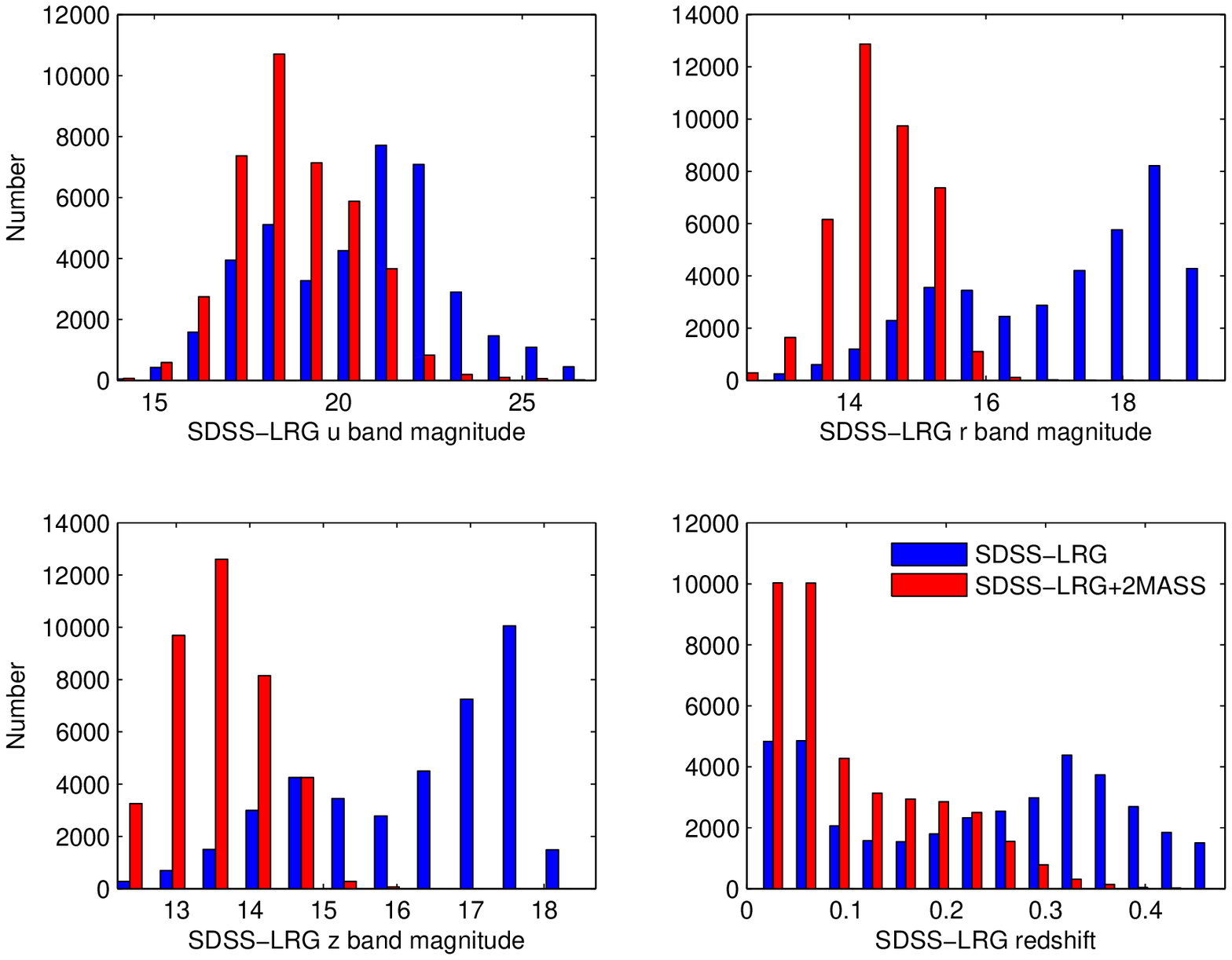}
\caption{Same as Figure~\ref{fig:Figure10} except we use Data Sets 2 (blue)
and 6 (red)}
\label{fig:Figure11}
\end{figure}

\subsection{Systematics}\label{sec:systematics}

\begin{figure}[Figure12]
\plotone{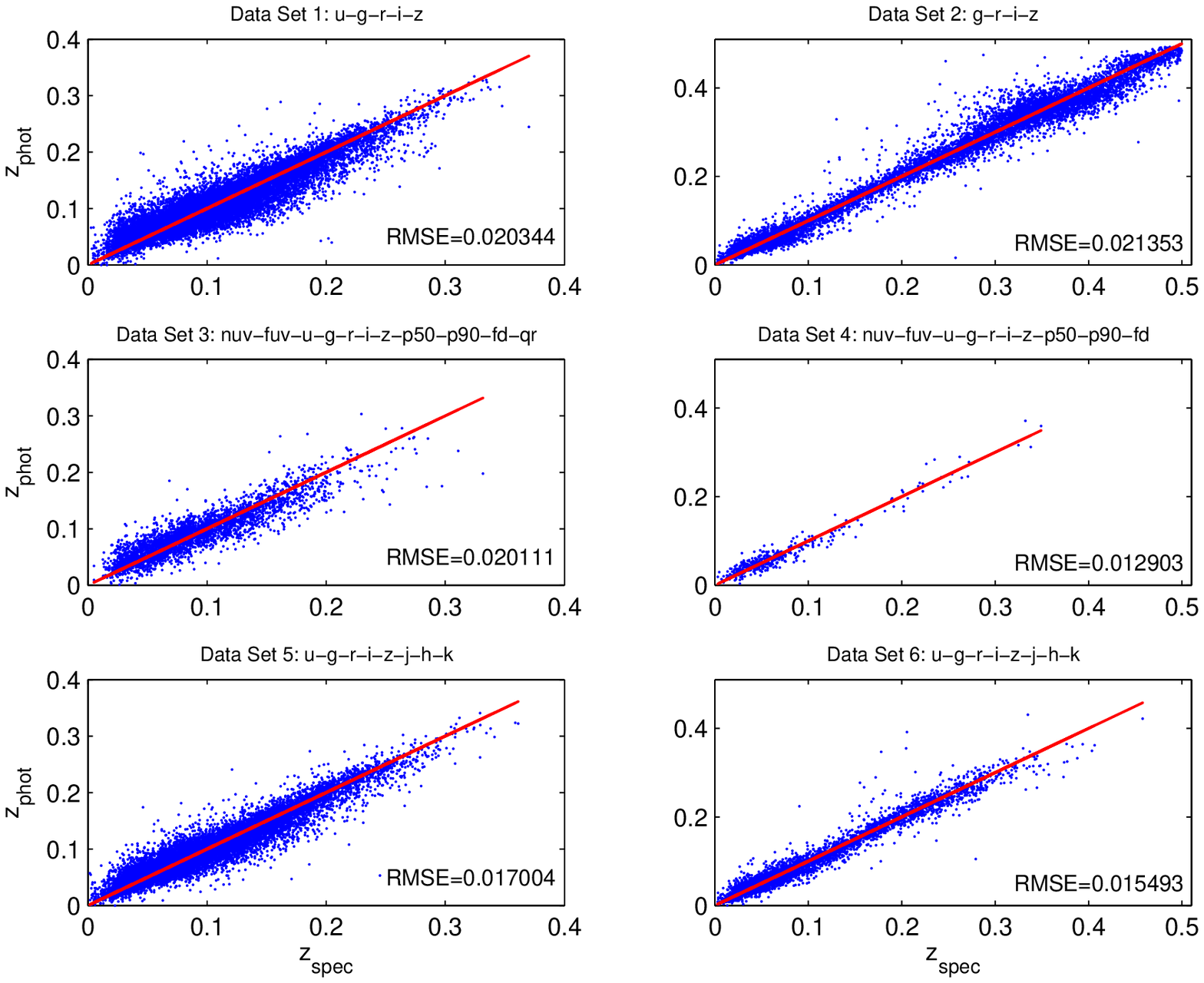}
\caption{Spectroscopic redshift plotted again predicted
photometric redshift for the best performing input from
each of the data sets in Table \ref{tbl-1}.} \label{fig:Figure12}
\end{figure}

\begin{figure}[Figure13]
\plotone{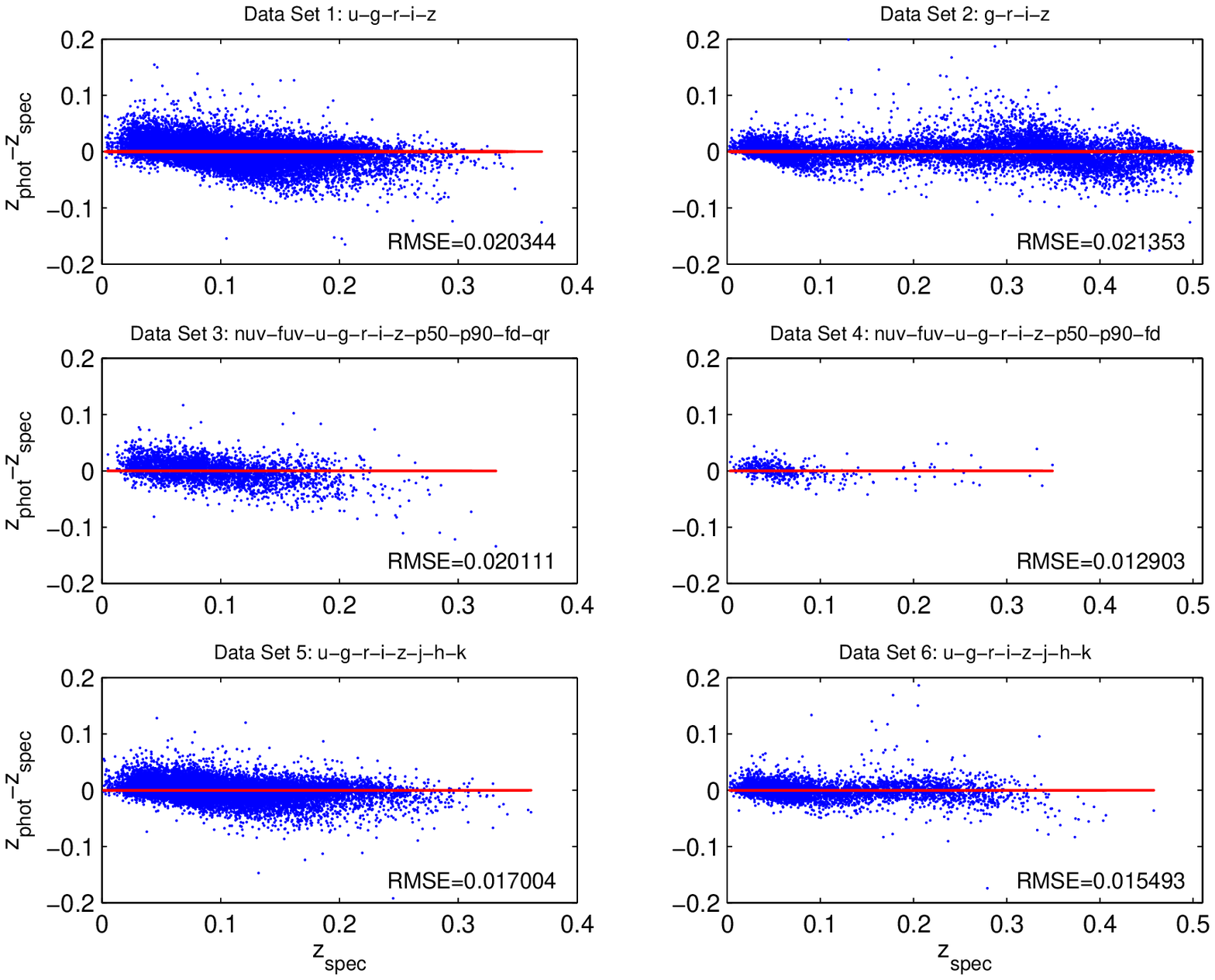}
\caption{Residuals as a function of spectroscopic redshift for the
best performing input from each of the Data Sets in Table \ref{tbl-1}.}
\label{fig:Figure13}
\end{figure}

In Figures \ref{fig:Figure12} and \ref{fig:Figure13}, we plot the redshifts and
residuals, respectively, for those data sets that yield the lowest RMSE.  The
actual RMSE is also indicated in each plot. There appears to be a systematic
shift above the regression line for redshifts less than 0.1 and below the
regression line between 0.1$<$z$<$0.2 for Data Sets 1, 3 and 5. This effect
has been seen or discussed in many papers on this 
topic \cite[e.g.][]{Collister04,DAbrusco07,Ball08,Wang09}.

At low redshifts (z$<$0.1) the bias in the regression line seen in
Figure \ref{fig:Figure12} (Data Set 1) is probably caused by the lack of deep
$u$-band data (see Figures \ref{fig:Figure8} and \ref{fig:Figure9}). When
supplemented by the GALEX data the bias looks to be slightly improved in Data
Set 3 (see Figures \ref{fig:Figure12} and \ref{fig:Figure13}). The bias seen
in between redshifts of 0.1$<$z$<$0.2 for the SDSS-MGS data sets
(Data Sets 1,3,5) is probably due to degeneracies in the spectral features of
those galaxies. This bias appears to be less with the addition of GALEX or
2MASS magnitudes, but it is still present nonetheless.

\subsection{Comparison with other work}\label{sec:comparisons}

In Paper I, we attempted to make comparisons between our
more primitive version of GPs (limited to 1000 training samples)
and several other well--known methods that we ran ourselves
(see Paper I, Tables 4--6) which included linear and quadratic
regression, the neural network ANNz package by \cite{Collister04},
and our own neural network type code called Ensemble Modeling (E-Model).
In Table~\ref{tbl-3}, we give the reader some appreciation of the abilities
of our updated GP method. We compare our new GP method with a representative
sample of recent work on two easily comparable data sets:
Data Set 1 using $u$-$g$-$r$-$i$-$z$ inputs and Data Set 2 using only
$u$-$g$-$r$-$i$-$z$ inputs.

\begin{deluxetable}{llll}
\tablecolumns{5}
\tablewidth{0pc}
\tablecaption{Photometric Redshift estimator comparisons for $u$-$g$-$r$-$i$-$z$ inputs\label{tbl-3}}
\tabletypesize{\scriptsize}
\tablehead{
\colhead{Method Name} &
\colhead{$\sigma_{rms}$\tablenotemark{a}} &
\colhead{Data Set\tablenotemark{b}} &
\colhead{Source} 
}
\startdata
CWW                       & 0.0666        & MGS SDSS-EDR & \cite{Csabai03} \\
Bruzual-Charlot           & 0.0552        & MGS SDSS-EDR & \cite{Csabai03} \\
ClassX                    & 0.0340        & MGS SDSS-DR2 & \cite{Suchkov05} \\
Polynomial                & 0.0318        & MGS SDSS-EDR & \cite{Csabai03} \\
Kd-tree                   & 0.0254        & MGS SDSS-EDR & \cite{Csabai03} \\
Support vector machine    & 0.0270        & MGS SDSS-DR2 & \cite{Wadadekar05} \\
Artificial neural network & 0.0229        & MGS SDSS-DR1 & \cite{Collister04} \\
Nearest neighbor          & 0.0207        & MGS SDSS-DR5 & \cite{Ball08} \\
                          & 0.0198        & MGS SDSS-DR5 & \cite{Ball08} \\
Hybrid Bayesian           & 0.0275        & MGS SDSS-DR5 & \cite{WG08} \\
Linear regression         & 0.0283 0.0282 0.0284 & MGS SDSS-DR3 & \cite{Way06}\\
Quadratic regression      & 0.0255 0.0255 0.0255 & MGS SDSS-DR3 & \cite{Way06}\\
ANNz\tablenotemark{c}     & 0.0206 0.0205 0.0208 & MGS SDSS-DR3 & \cite{Way06}\\
Ensemble model            & 0.0201 0.0198 0.0205 & MGS SDSS-DR3 & \cite{Way06}\\
Gaussian process 1000\tablenotemark{d}& 0.0227 0.0225 0.0230  & MGS SDSS-DR3 & \cite{Way06}\\
Gaussian process\tablenotemark{e}&0.0201 0.0200 0.0201& MGS SDSS-DR3 & This work: Data Set 1\\
\hline
\hline
Nearest neighbor          & 0.0243        & LRG SDSS-DR5 & \cite{Ball08} \\
                          & 0.0223        & LRG SDSS-DR5 & \cite{Ball08} \\
Hybrid                    & 0.0300        & LRG SDSS-DR3 & \cite{Padmanabhan05} \\
Linear regression\tablenotemark{f}& 0.0289 0.0289 0.0289 & LRG SDSS-DR5 & This work: Data Set 2\\
Quadratic regression\tablenotemark{f}& 0.0240 0.0240 0.0240 & LRG SDSS-DR5 & This work: Data Set 2\\
ANNz\tablenotemark{c}     & 0.0207 0.0205 0.0210 & LRG SDSS-DR5 & This work: Data Set 2\\
Ensemble Model\tablenotemark{f}& 0.0221 0.0220 0.0221 & LRG SDSS-DR5 & This work: Data Set 2\\
Gaussian Process\tablenotemark{e}& 0.0220 0.0217 0.0240 & LRG SDSS-DR5 & This work: Data Set 2\\
\enddata
\tablenotetext{a}{The $\sigma_{rms}$ cited here are for rough comparison only.
No error bounds are included for the cited publications since many
do not give error bounds or they are not handled in a consistent fashion
across publications. For this paper's results, we quote the
bootstrapped 50\%, 10\%, and 90\% confidence levels as in Paper I.}
\tablenotetext{b}{MGS: Main Galaxy sample, LRG = Luminous Red Galaxy sample,
SDSS-EDR = SDSS Early Data Release \citep{Stoughton02}, SDSS-DR1 = SDSS Data
Release One \citep{Abazajian03}, SDSS-DR2 = SDSS Data Release Two \citep{Abazajian04},
SDSS-DR3 = SDSS Data Release Three \citep{Abazajian05},
SDSS-DR5 = SDSS Data Release Five \citep{AM07}.}
\tablenotetext{c}{Uses the ANNz code of \citep{Collister04}.}
\tablenotetext{d}{GP algorithm limited to 1000 training samples.}
\tablenotetext{e}{GP algorithm SR-VP with 80,000 training samples
and rank=800.}
\tablenotetext{f}{See Paper I \citep{Way06} for details on these algorithms.}
\end{deluxetable}

\section{Conclusion}\label{sec:section7}

We have demonstrated that with new non-sparse matrix inversion techniques
and a better choice of kernel (or transfer function if you prefer) that
GPR is a competitive way to obtain accurate photometric
redshifts for low-redshift surveys such as the SDSS.  However, several caveats
must be noted regarding the estimation of photometric redshifts from combined
catalogs of the SDSS and 2MASS as well as the SDSS and GALEX as discussed in
Section \ref{sec:section6}.

The SDSS + 2MASS and SDSS + GALEX cross-match results are astoundingly good
in some cases, but this occurs even when the only bandpasses used are the
$u$-$g$-$r$-$i$-$z$ of the SDSS cross-matched set. This is clearly a case
where we are sampling a smaller range of redshifts and magnitudes,
which makes the regression job easier regardless of the algorithm.  This shows
that one has to be careful when quoting ``better" results from a cross-match 
of any catalog.

We also demonstrate that the addition of many SDSS morphological parameters
does not systematically improve our regression results.  For a low--redshift
survey like the SDSS, it makes intuitive sense that the Petrosian radii
would help given the angular--diameter--distance relation, but that does not
appear to be the case here unlike that of other studies
\cite[e.g.,][]{Wadadekar05}.

%%%%%%%%%%%%%%%%%%%%%%%%%%%%%%%%%%%%%%%%%%%%%%%%%%%%%%%%%%%%%%%%%%%%%%%

\acknowledgements

The papers associated with this project and the code used to generate
the results from this paper are available on the NASA Ames Dashlink Web site
https://dashlink.arc.nasa.gov/algorithm/stablegp

M.J.W thanks Jim Gray, Ani Thakar, Maria SanSebastien, and Alex Szalay for their
help in cross-matching the catalogs used herein. Thanks goes to the
Astronomy Department at Uppsala University in Sweden for their generous
hospitality while part of this work was completed.  M.J.W. acknowledges funding
received from the NASA Applied Information Systems Research Program.
A.N.S. thanks the NASA Aviation Safety Integrated Vehicle Health Management
project for support in developing the GP-V method.
The authors would like to acknowledge support for this project from the Woodward
Fund, Department of Mathematics, San Jose State University.
The authors acknowledge support from the NASA Ames Research Center
Director's Discretionary Fund.

Funding for the SDSS has been provided by
the Alfred P. Sloan Foundation, the Participating Institutions, the National
Aeronautics and Space Administration, the National Science Foundation,
the U.S. Department of Energy, the Japanese Monbukagakusho, and the Max
Planck Society. The SDSS Web site is http://www.sdss.org/.

The SDSS is managed by the Astrophysical Research Consortium for
the Participating Institutions. The Participating Institutions are the
University of Chicago, Fermilab, the Institute for Advanced Study, the
Japan Participation Group, The Johns Hopkins University, Los Alamos National
Laboratory, the Max-Planck-Institute for Astronomy, the
Max-Planck-Institute for Astrophysics, New Mexico State University,
University of Pittsburgh, Princeton University, the United States Naval
Observatory, and the University of Washington.

This publication makes use of data products from the Two Micron All Sky Survey,
which is a joint project of the University of Massachusetts and the Infrared
Processing and Analysis Center/California Institute of Technology, funded by
the National Aeronautics and Space Administration and the National Science
Foundation.

The Galaxy Evolution Explorer (GALEX) is a NASA Small Explorer.
The mission was developed in cooperation with the Centre
National d'\'{E}tudes Spatiales of France and the Korean Ministry of
Science and Technology.

This research has made use of NASA's Astrophysics Data System Bibliographic
Services.

%%%%%%%%%%%%%%%%%%%%%%%%%%%%%%%%%%%%%%%%%%%%%%%%%%%%%%%%%%%%%%%%%%%%%%%


\begin{thebibliography}{}

\bibitem[Abazajian et al.(2003)]{Abazajian03}
Abazajian, K., et al. 2003, \aj, 126, 2081

\bibitem[Abazajian et al.(2004)]{Abazajian04}
Abazajian, K., et al. 2004, \aj, 128, 502

\bibitem[Abazajian et al.(2005)]{Abazajian05}
Abazajian, K., et al. 2005, \aj, 129, 1755

\bibitem[Adelman-McCarthy et al.(2007)]{AM07}
Adelman-McCarthy, J.K. et al. 2007, \apjs, 172, 634 

\bibitem[Ball et al.(2008)]{Ball08}
Ball, N.M., Brunner, R.J., Myers A.D., Strand, N.E., Alberts, S.L., \&
Tcheng, D. 2008, \apj, 683, 12

\bibitem[Bernstein \& Huterer(2009)]{BH2009}
Bernstein, G. \& Huterer, D. 2009, arXiv:0909.2782v1

\bibitem[Blanton et al.(2005)]{Blanton05}
Blanton, M.R., et al 2005, \aj, 129, 2562

\bibitem[Carliles et al.(2008)]{Carliles08}
Carliles, S., et al. 2008, in ASP Conf. Ser. 394,
Astronomical Data Analysis Software and Systems, ed. R.W. Argyle, P.S.
Bunclark, \& J.R. Lewis (San Francisco, CA: ASP), 521

\bibitem[Collister \& Lahav(2004)]{Collister04}
Collister, A.A. \& Lahav, O. 2004, \pasp, 116, 345

\bibitem[Csabai et al.(2003)]{Csabai03}
Csabai, I., et al. 2003, \aj, 125, 580

\bibitem[D'Abrusco et al.(2007)]{DAbrusco07}
D'Abrusco, R., Staiano, A., Giuseppe, L., Brescia, M., Paolillo, M. De Filippis,
E. \& Tagliaferri, R. 2007, \apj, 663, 752

\bibitem[Efron \& Tibshirani(1993)]{ET93}
Efron, B., \& Tibshirani, R.J. 1993, An introduction to the
bootstrap. New York, Chapman \& Hall

\bibitem[Eisenstein et al.(2001)]{Eisenstein01}
Eisenstein et al. 2001, \aj, 122, 2267

\bibitem[Fine \& Scheinberg(2001)]{fine2001}
Fine, S. \& Scheinberg, K. 2001, J. Mach. Learn. Res. 2, 243

\bibitem[Foster et al.(2009)]{foster09}
Foster, L., et al. 2009 J. Mach. Lear. Res. 10, 857
Research, 10, 857

\bibitem[Giavalisco et al.(2004)]{Giavalisco04}
Giavalisco, M. et al. 2004 \apj, 600, L93

\bibitem[Golub \& Van Loan(1996)]{golub1996}
Golub, G.H. \& Van Loan, C.F. 1996, Matrix Computations (3rd ed.; 
Baltimore, MD, USA: Johns Hopkins Univ. Press)

\bibitem[Ivezic et al.(2008)]{Ivezic08}
Ivezic, Z., Tyson, J.A., Allsman, R., Andrew, J., Angel, R., et al 2008,
arXiv:0805.2366v1 

\bibitem[Kaczmarczik et al.(2009)]{kaczmarcik2009}
Kaczmarczik, M.C., Richards, G.T., Mehta, S.S. \& Schlegel, D.J. 2009, \aj,
138, 19

\bibitem[Kaiser et al.(2002)]{Kaiser02}
Kaiser, N. et al. 2002, Proc. SPIE, 4836, 154

\bibitem[Kurtz et al.(2007)]{Kurtz07}
Kurtz, M.J., et al. 2007, \aj, 134, 1360

\bibitem[Le F\`{e}vre et al.(2004)]{LeFevre04}
Le F\`{e}vre O. et al. 2004, A\&A, 417, 839

\bibitem[Martin et al.(2005)]{Martin05}
Martin, D. C., et al. 2005, \apj, 619, L1

\bibitem[Neal(1996)]{Neal1996}
Neal R.M. 1996, Bayesian Learning for Neural Networks (New York: Springer)

\bibitem[Padmanabhan et al.(2005)]{Padmanabhan05}
Padmanabhan, N. et al. 2005, \mnras, 359, 327

\bibitem[Petrosian(1976)]{Petrosian76}
Petrosian, V. 1976, \apj, 209, L1

\bibitem[Poggio \& Giroso(1990)]{PG1990}
Poggio, T., \& Girosi, F. 1990 Proc. IEEE, 78, 1481.

\bibitem[Rasmussen \& Williams(2006)]{RW2006}
Rasmussen, C.E. \& Williams, C.K.I. 2006 Gaussian Processes
for Machine Learning (Cambridge, MA: MIT Press)

\bibitem[Seeger et al.(2003)]{seeger2003}
Seeger, M., Williams, C., \& Lawrence, N.D. 2003, in Proc. 9th Int.
Workshop on Artificial Intelligence and Statistics, Fast Forward Selection
to Speed Up Sparse Gaussian Process Regression, ed. C.M. Bishop \& B.J. Frey
(San Francisco, CA: Morgan Kaufmann)

\bibitem[S\'{e}rsic(1968)]{Sersic68}
S\'{e}rsic, J.L. 1968, Atlas de Galaxias Australes (Cordoba, Argentina:
Observatorio Astronomico)

\bibitem[Skrutskie et al.(2006)]{Skrutskie06}
Skrutskie, M.F. et al. 2006, \aj, 131, 1163

\bibitem[Stabenau et al.(2008)]{Stabenau08}
Stabenau, H.F., Connolly, A., \& Bhuvnesh, J. 2008, \mnras, 387, 1215

%%% SDSS EDR
\bibitem[Stoughton et al.(2002)]{Stoughton02}
Stoughton, C. et al. 2002, \aj, 123, 485

\bibitem[Strauss et al.(2002)]{Strauss02}
Strauss, M.A., et al. 2002, \aj, 124, 1810

\bibitem[Suchkov et al.(2005)]{Suchkov05}
Suchkov, A.A., Hanisch, R.J., \& Margon, B. 2005, \aj, 130, 2439

\bibitem[Wadadekar(2005)]{Wadadekar05}
Wadadekar, Y. 2005, \pasp, 117, 79

\bibitem[Wang et al.(2008)]{Wang08}
Wang, D., Zhang, Y., Liu, C. \& Zhao, Y. 2008, ChJAA, 8, 119

\bibitem[Wang et al.(2009)]{Wang09}
Wang, T., Huang, J. \& Gu, Q. 2009, Res. Astron. Astrophys., 9, 390

\bibitem[Way \& Srivastava(2006)]{Way06}
Way, M.J. \& Srivastava, A.N. 2006, \apj, 647, 102

\bibitem[Whaba(1990)]{wahba1990}
Wahba, G. 1990, Spline Models for Observational Data (Philadelphia, PA: SIAM)

\bibitem[Wray \& Gunn(2008)]{WG08}
Wray, J.J. \& Gunn, J.E. 2008, \apj, 678, 144

\bibitem[York et al.(2000)]{York2000}
York, D.G., et al. 2000, \aj, 120, 1579

\end{thebibliography}
\end{document}